\documentclass[]{iopart}

\expandafter\let\csname equation*\endcsname\relax
\expandafter\let\csname endequation*\endcsname\relax

\usepackage{graphicx}   
\usepackage{epstopdf}           
\usepackage{cite}
\usepackage[perpage,marginal]{footmisc}
\usepackage{etoolbox}
\usepackage{marvosym}
\usepackage{upgreek}
\usepackage{pdflscape}
\usepackage[unicode]{hyperref}
\usepackage[all]{hypcap}
\usepackage{amsmath,amsfonts,amssymb}
\usepackage{color}
\usepackage[usenames,dvipsnames]{xcolor}
\usepackage{bm}
\usepackage{longtable}
\usepackage[T1]{fontenc}
\usepackage[utf8]{inputenc}
\usepackage{verbatim}
\usepackage{pifont}
\usepackage{yfonts}
\usepackage{url}
\usepackage{multirow}
\usepackage{tensor}
\usepackage{mathtools}
\makeatletter
\def\url@leostyle{%
  \@ifundefined{selectfont}{\def\UrlFont{\sf}}{\def\UrlFont{\small\ttfamily}}}
\makeatother
\def\apj{Astrophysical Journal }                

\def\radcurv{\mathfrak{r}} 
\def\massdens{{\rho}}
\def\bareG{{G_{\star}}}
\def\ssA{{\scriptscriptstyle A}}
\def\ssB{{\scriptscriptstyle B}}

\newcommand{\gothr}{\textswab{r}}
\newcommand{\vp}{\varphi}
\newcommand{\cvp}{\bar{\varphi}}

\newcommand{\nn}{\nonumber}

\newcommand{\be}{\begin{equation}}
\newcommand{\ee}{\end{equation}}
\newcommand{\ba}{\begin{eqnarray}}
\newcommand{\eea}{\end{eqnarray}}
\newcommand{\bea}{\begin{eqnarray}}
\newcommand{\ea}{\end{eqnarray}}

\definecolor{darkgreen}{rgb}{0.2,0.7,0.2}

\makeatletter
\def\@mkboth#1#2{}
\newlength\appendixwidth
\preto\appendix{\addtocontents{toc}{\protect\patchl@section}}
\newcommand{\patchl@section}{%
  \settowidth{\appendixwidth}{\textbf{Appendix }}%
  \addtolength{\appendixwidth}{1.5em}%
  \patchcmd{\l@section}{1.5em}{\appendixwidth}{}{\ddt}%
}
\makeatother

\def\n{\nu}

\def\g{\gamma}
\def\l{\lambda}

\def\nn{\nonumber}
\def\ii{{\rm i}}

\usepackage[normalem]{ulem}
\usepackage{fancyhdr}

\makeatletter
\renewcommand\footnoterule{%
  \kern-3\p@
  \hrule\@width2.5cm
  \kern2.6\p@}
\makeatother

\numberwithin{equation}{section}
\numberwithin{figure}{section}

\newcommand{\sAE}{\ifmmode\text{\AE}\else\AE\fi}

\makeatletter
\renewcommand\@appendixstar{\@@par
 \ifnumbysec
 \@addtoreset{table}{section}
 \@addtoreset{figure}{section}\fi
 \setcounter{section}{0}
 \setcounter{subsection}{0}
 \setcounter{subsubsection}{0}
 \setcounter{equation}{0}
 \setcounter{figure}{0}
 \setcounter{table}{0}
  \def\thesection{Appendix \Alph{section}}
   \def\thesubsection{\Alph{subsection}}
 \def\theequation{\ifnumbysec
      \Alph{section}.\arabic{equation}\else
      \Alph{section}\arabic{equation}\fi}
 \def\thetable{\ifnumbysec
      \Alph{section}\arabic{table}\else
      A\arabic{table}\fi}
 \def\thefigure{\ifnumbysec
      \Alph{section}\arabic{figure}\else
      A\arabic{figure}\fi}}
\makeatother

\begin{document}



\pagestyle{fancy}
\lhead{Tensor-multi-scalar theories}
\chead{}
\rhead{\thepage}
\lfoot{}
\cfoot{}
\rfoot{}

\begin{center}
\title{Tensor-multi-scalar theories: relativistic stars and 3+1 decomposition}
\end{center}

\author{%
Michael~Horbatsch$^{1,2}$,
Hector~O.~Silva$^{2}$ ,
Davide~Gerosa$^{3}$,
Paolo~Pani$^{4,5}$,
Emanuele~Berti$^{2,5}$,
Leonardo~Gualtieri$^{4}$
and
Ulrich~Sperhake$^{2,3,6}$
}

\address{$^{1}$School of Mathematical Sciences,
University of Nottingham, Nottingham, NG7 2RD, UK}
\address{$^{2}$~Department of Physics and Astronomy,
The University of Mississippi, University, MS 38677-1848, USA}
\address{$^{3}$~Department of Applied Mathematics and Theoretical Physics, Centre for Mathematical Sciences, University of Cambridge,
Wilberforce Road, Cambridge CB3 0WA, UK}
\address{$^{4}$~Dipartimento di Fisica, ``Sapienza'' Universit\`a di Roma
\& Sezione INFN Roma 1, P.le A. Moro 2, 00185 Roma, Italy}
\address{$^{5}$~CENTRA, Departamento de F\'isica, Instituto Superior
T\'ecnico, Universidade de Lisboa, Avenida Rovisco Pais 1,
1049 Lisboa, Portugal}
\address{$^{6}$~Theoretical Astrophysics 350-17,
California Institute of Technology, Pasadena, CA 91125, USA}

\ead{michael.horbatsch@nottingham.ac.uk}

\begin{abstract}
  Gravitational theories with multiple scalar fields coupled to the
  metric and each other --- a natural extension of the well studied
  single-scalar-tensor theories --- are interesting phenomenological
  frameworks to describe deviations from general relativity in the
  strong-field regime. In these theories, the $N$-tuple of scalar
  fields takes values in a coordinate patch of an $N$-dimensional
  Riemannian target-space manifold whose properties are poorly
  constrained by weak-field observations. Here we introduce for
  simplicity a non-trivial model with two scalar fields and a
  maximally symmetric target-space manifold.  Within this model we
  present a preliminary investigation of spontaneous scalarization for
  relativistic, perfect fluid stellar models in spherical symmetry.
  We find that the scalarization threshold is determined by the
  eigenvalues of a symmetric scalar-matter coupling matrix, and that
  the properties of strongly scalarized stellar configurations
  additionally depend on the target-space curvature radius.
  In preparation for numerical relativity simulations, we also write
  down the $3+1$ decomposition of the field equations for generic
  tensor-multi-scalar theories.
\end{abstract}


\clearpage
{\tableofcontents}
\clearpage
\section{Introduction}
\label{sec:intro}

Modifications of general relativity (GR) often lead to the
introduction of additional degrees of
freedom~\cite{Berti:2015itd}. The simplest and best studied extension
of GR is scalar-tensor (ST) theory, in which one or more scalar fields
are included in the gravitational sector of the action through a
non-minimal coupling between the Ricci scalar and a function of the
scalar field(s). A further motivation to study ST theories is that
they appear in different contexts in high-energy physics: they can be
obtained as the low-energy limit of string theories
~\cite{polchinski1998string}, in Kaluza-Klein-like
models~\cite{Duff:1994tn} or in braneworld
scenarios~\cite{Randall:1999ee,Randall:1999vf}. Moreover, ST theories
play an important role in cosmology~\cite{Clifton:2011jh}.

Almost 60 years ago, in an attempt to implement Mach's ideas in a
relativistic theory of gravity, Jordan, Fierz, Brans and Dicke
proposed a specific ST theory (commonly referred to as ``Brans-Dicke
theory'') as a possible modification of
GR~\cite{Jordan:1959eg,Fierz:1956zz,Brans:1961sx}. Their theory is
still viable, but it has since been constrained to be extremely close
to GR by Solar System and binary pulsar observations
\cite{Will:2005va}.  Brans-Dicke theory was generalized by Bergmann
and Wagoner, who considered the most general ST theory with a single
scalar field and an action at most quadratic in derivatives of the
fields~\cite{Bergmann:1968ve,Wagoner:1970vr}. In 1992 Damour and
Esposito-Far\`ese introduced and investigated tensor-multi-scalar
(TMS) theory, a generalization of the Bergmann-Wagoner theory to an
arbitrary number of scalar fields~\cite{Damour:1992we}. 
Multiple scalar degrees of freedom are a generic prediction of string
theories and theories involving extra dimensions (see
e.g.~\cite{Kainulainen:2004vk,Albrecht:2001xt,Damour:1994ya,Damour:1994zq}).
In recent years, Bergmann-Wagoner theory has been extensively studied
in the case of a single scalar field (see
e.g.~\cite{Chiba:1997ms,2003sttg.book.....F,Horbatsch:2010hj,Sotiriou:2015lxa}
and references therein). In comparison, very limited attention has
been devoted to the phenomenological implications of TMS theory.

Even in the simplest case of a single scalar, ST theories give rise to
interesting phenomenology. Although their action is linear in the
curvature tensor, and scalar-matter couplings are highly constrained
by observational bounds from the Solar System~\cite{Bertotti:2003rm},
ST theories can modify the strong-field regime of GR. Indeed, the
equations of structure describing compact stars can admit
non-perturbative solutions where the scalar fields can have large
amplitudes~\cite{Damour:1993hw}. This phenomenon, called {\it
  spontaneous scalarization}, can significantly affect the masses and
radii of neutron stars.  Spontaneous scalarization is strongly
constrained by binary pulsar observations~\cite{Freire:2012mg}, but it
could still leave a signature in the late inspiral of compact binaries
through the so-called ``dynamical
scalarization''~\cite{Barausse:2012da,Palenzuela:2013hsa,Shibata:2013pra},
which may be observable by advanced gravitational-wave
detectors~\cite{Sampson:2014qqa,Taniguchi:2014fqa}.

In this article we study the phenomenology of TMS theories.
The importance of our work for black hole physics is to a significant
extent of indirect nature. Stellar collapse represents, of course, one
of the most important channels for black hole formation, and ST
theories of gravity are more likely to produce experimental signatures
during black hole formation than in the dynamics of the remnant
``quiescent'' black hole spacetimes \cite{Novak:1997hw,Harada:1996wt}.
%
%
The reason is that there are strong no-hair theorems\footnote{Hairy
  black hole solutions are possible in presence of a complex scalar
  field (i.e. two real scalar fields), as long as its phase is
  time-dependent~\cite{Herdeiro:2014goa}. No-hair theorems can also be
  evaded if the potential is non-convex; in this case the solutions
  are unstable, but their growth time can be extremely
  large~\cite{Herdeiro:2015waa}.}  implying that stationary, vacuum
black hole solutions in ST and TMS theories must be identical to GR
(cf.~\cite{Damour:1992we,Bekenstein:1995un,Sotiriou:2011dz,Heusler:1995qj,Chrusciel:2012jk,HeuslerBook}
and Sotiriou's contribution to the present
volume~\cite{Sotiriou:2015pka}).
In addition, it has been shown that the dynamics of black-hole
binaries in ST theory is undistinguishable from that in GR up to $2.5$
post-Newtonian order~\cite{Will:1989sk,Mirshekari:2013vb} or -- in the
case of extreme mass-ratio -- to all post-Newtonian
orders~\cite{Yunes:2011aa}. This result does not apply in the presence
of non-trivial boundary
conditions~\cite{Horbatsch:2011ye,Healy:2011ef,Berti:2013gfa}, but in
this case violations of the no-hair theorem would most likely be
unobservable.  Nevertheless, black hole binaries have been studied in
the framework of single-scalar theories using numerical relativity
(cf.~\cite{Healy:2011ef,Berti:2013gfa}), and the extension of such
studies to TMS theories may still reveal surprising discoveries.

In order to pave the way for numerical investigations of black holes
and neutron stars in TMS theories, in Section~\ref{sec:3p1formulation}
we write down the TMS field equations in a 3+1 formalism that is
suitable for numerical evolutions of compact binary systems.


A promising avenue to understand the experimental implications of
TMS theories (following the reasoning of
\cite{Damour:1993hw,Damour:1996ke}) is to focus first on the coupling
between the various scalars and matter.
The scalar fields take values in a coordinate patch of a Riemannian
{\it target-space} manifold.  It is natural to ask whether this
additional geometric structure
can leave a detectable signature in compact stars and/or in the late
inspiral of compact binaries, while still allowing for solutions
compatible with binary pulsar observations.

This paper is a preliminary investigation in this direction. We will
mainly focus on a simple non-trivial TMS model with two scalar fields
and a vanishing potential. This model contains the main novel features
that distinguish TMS theories from single-scalar theories, i.e. the
presence of a target-space manifold with non-trivial Riemannian
structure and non-vanishing curvature, and the presence of a
continuous symmetry that is spontaneously broken by scalarized
solutions.  A more systematic study of scalarization in this prototype
TMS theory will be the topic of a future publication, and it should
give us useful insight into strong-field effects that characterize
more general TMS theories.

The plan of the paper is as follows. In Section~\ref{sec:theory} we
introduce the action and the field equations of TMS theory. We then
specialize the field equations to the case of two scalar fields with
vanishing potential and a maximally symmetric target-space manifold.
In Section~\ref{sec:stellarstructure} we derive the equations for
slowly rotating relativistic stars, and we perform numerical
integrations of these equations in the non-rotating case. In
Section~\ref{sec:3p1formulation} we derive the $3+1$ decomposition of
the TMS field equations, which can be used to perform fully non-linear
numerical evolutions. In Section~\ref{sec:conclusions} we draw some
conclusions and point out possible directions for future work.
The Appendices contain some technical material on the structure of
target spaces in our model (\ref{app:targetspaces}), experimental
constraints (\ref{app:constraints}) and perturbative arguments to
predict the spontaneous scalarization threshold in models with two
scalar fields (\ref{app:linear}).

\section{Tensor-multi-scalar theories: action, field equations, scalar-matter couplings, and symmetries}
\label{sec:theory}

\subsection{Units, notation and conventions}
\label{sec:units}
Throughout the paper we use units with $c=1$. The gravitational
constant measured in a Cavendish experiment is denoted by $G$, while
the ``bare'' gravitational constant appearing in the action is denoted
by $G_\star$: the relation between the two constants is written down
explicitly in Eq.~(\ref{eq:GGstar}).  Indices on space-time tensors
are denoted by Greek letters and take values $0 , \ldots , 3$, and
space-time coordinates are denoted by $x^{\mu}$.
The Lorentzian space-time metric is taken to have signature $(- , + , + , +)$, and its components are denoted by $g_{\mu
  \nu}(x)$.  The conventions for the Riemann curvature tensor and its
contractions, as well as the notation for symmetrization and
anti-symmetrization of tensors, are those of Misner, Thorne, and Wheeler~\cite{1973grav.book.....M}.

The $N$-tuple of scalar fields
$\vp^A(x) = (\vp^{1}(x), \ldots , \vp^{N}(x))$ takes values in a
coordinate patch of an $N$-dimensional Riemannian target-space
manifold.  Indices on target-space tensors are denoted by early
capital Roman letters $A,\,B,\,C,\,\ldots$, and take integer values
$1, \ldots , N$. Components of the Riemannian target-space metric are
denoted by $\gamma_{AB}(\vp)$, and the associated Christoffel symbols
are denoted by $\gamma^{C}_{\phantom{C}AB}(\vp)$. The target-space
Riemann curvature tensor is denoted by
$\mathcal{R}^{A}_{\phantom{A}BCD}(\vp)$, with obvious notation for
derived quantities such as the Ricci tensor and the Ricci scalar.  If
the target space has a Hermitian structure\footnote{For an
  introduction to Hermitian structures and complex differential
  geometry, see e.g.~\cite{moroianu_book}.}, then indices on
complexified tensors are denoted by lower-case Roman letters, and take
values $1 , 2 , \ldots , N/2$. Holomorphic coordinates are denoted by
$(\varphi^{a},\bar{\varphi}^{a})$, and the components of the Hermitian
metric in these coordinates are denoted by
$\gamma_{\bar{a}b}(\varphi,\bar{\varphi})$.
For reference, in Table~\ref{default} we provide an overview of the
meaning of the various symbols and conventions used in this paper.

\begin{table}[]
\begin{center}
\begin{tabular}{cc}
\hline\\[-10pt] \hline  \\[-8pt]
$G$  & Gravitational constant from a Cavendish experiment\\
$G_\star$ & Bare gravitational constants appearing in the action\\
$\mu,\nu,\rho$ & Spacetime indices \\
$x^\mu$ & Spacetime coordinates \\
$g_{\mu\nu}$ & Spacetime metric in the Einstein frame \\
$\nabla_{\mu}$ & Covariant derivative associated with $g_{\mu\nu}$ \\
$R^\mu{}_{\nu\rho\sigma}$ & Spacetime Riemann tensor \\
$A,B,C$ & Scalar-field indices in real notation \\
$N$ & Number of  scalar fields\\
$\varphi^{A}$ &  Gravitational scalar fields in real notation \\
$\gamma_{AB}$ &  Target-space  metric in real notation \\
$\gamma^A_{BC}$ & Christoffel symbols on the target space in real notation \\
$\mathcal{R}^A{}_{BCD}$ & Target-space  Riemann tensor \\
$\Psi$ & Non-gravitational fields \\
$\tilde{g}_{\mu \nu}$ & Spacetime metric in the Jordan frame \\
$A(\varphi)$ & Einstein-Jordan frame conformal factor \\
$V(\varphi)$ & Scalar-field potential \\
$a,b,c$ & Scalar-field indices in complex notation \\
$\varphi^{a},\bar\varphi^{a}$ &  Gravitational scalar fields in complex notation  \\
$\gamma_{{\bar a}b}$ &  Target-space  metric in complex notation \\
$\gamma^a_{\bar{b}c},~\gamma^a_{bc}$ & Christoffel symbols on the target space
in complex notation \\
$\gothr$ & Target-space curvature radius ($N=2$)\\
$\kappa(\varphi,\bar\varphi)$ & Scalar-matter coupling function
                                ($N=2$), see Eq.~(\ref{eq:kappa_def})\\
$\alpha^\ast,\bar\alpha^\ast$ & Linear-term coefficients in the expansion of $\log A(\psi,\bar\psi)$\\
$\beta_0,\beta^\ast_1,\bar\beta^\ast_1$ & Quadratic-term coefficients in the expansion of $\log A(\psi,\bar\psi)$ \\
$\theta$ & Generic rotation angle in the target-space complex plane \\
$\alpha,\beta_1$ &  Redefinition of $\alpha^\ast,\beta^\ast_1,$ after rotation\\
$\psi,\bar\psi$ & Redefinition of the fields $\varphi,\bar\varphi$ after rotation\\
$Z$ & ${\rm Re}[\psi]$\\
$W$ & ${\rm Im}[\psi]$\\
\hline\\[-10pt]
$t,r,\theta,\phi$ &  Spacetime coordinates for stellar models  \\
$\nu(r)$ &  Lapse function \\
$m(r)=r \mu(r)$ &  Mass function \\
$\omega(r)$ &  Fluid differential angular velocity \\
$\Omega$ &  Angular velocity of the star \\
$\rho$ & Fluid mass-energy density\\
$P$ & Fluid pressure\\
$n_B$ & Baryon density\\
$u^\mu,\tilde u^\mu$ & Fluid 4-velocity in the Einstein/Jordan frame\\
Subscript ``$0$'' & Previous quantities evaluated at the star's center, $r=0$\\
$R,\tilde R$ & Stellar radius in the Einstein/Jordan frame\\
$\psi_\infty$ & Asymptotic value of the scalar field\\
$M$ & Gravitational mass of the star\\
$Q$ & Scalar charge of the star\\
$M_B$ & Baryonic mass of the star\\
$K,n_0,m_b,\gamma$ & Equation of state parameters, see Eq.~(\ref{eq:EOS_rho})

\end{tabular}
\end{center}
\caption{Variables and conventions used in this paper. Quantities above the horizontal line define the theory; quantities below the horizontal line refer to stellar models.}
\label{default}
\end{table}%

\subsection{Action and field equations for $N$ real scalars}
\label{sec:action_fieldeq}

We consider a gravitational theory with metric tensor
$g_{\mu \nu}$, and scalar fields $\vp^{1}, \ldots , \vp^{N}$
which take values in a coordinate patch of an
$N$-dimensional target-space manifold.  We assume that all
non-gravitational fields, denoted collectively by $\Psi$, couple only
to the Jordan-frame metric $\tilde{g}_{\mu \nu}=A^{2}(\vp)g_{\mu\nu}$,
so that the matter action has the functional form
$S_{\rm m}[\Psi ; \tilde{g}_{\mu \nu} ]$.  This assumption guarantees
that the Weak Equivalence Principle (WEP), which has been
experimentally verified with great accuracy~\cite{Will:2005va}, will
hold. The quantity $A(\vp)$ is a conformal factor relating the metrics
$\tilde{g}_{\mu \nu}$ and $g_{\mu\nu}$.

The most general action which is invariant under space-time and
target-space diffeomorphisms (up to boundary terms and field
redefinitions), and has at most two space-time derivatives, can be
written in the form~\cite{Damour:1992we}
\begin{align}
S = \frac{1}{4 \pi G_{\star}} \int d^{4}x \sqrt{-g}
\left[ \frac{R}{4} - \frac{1}{2}g^{\mu \nu} \gamma_{AB}(\vp) \partial_{\mu}\vp^{A} \, \partial_{\nu}\vp^{B} - V(\vp) \right]
+ S_{\rm m}[A^{2}(\vp)g_{\mu \nu} ; \Psi] \,, \nonumber \\
\label{eq:action_ef}
\end{align}
where $G_{\star}$ is a bare gravitational constant, and $g$ and
$R$ are the determinant and Ricci scalar of $g_{\mu \nu}$,
respectively.  The positive-definiteness of the target-space
Riemannian metric $\gamma_{AB}(\vp)$ guarantees the absence of
negative-energy excitations.  The scalars $\vp^{A}$ are 
dimensionless and the potential $V(\vp)$ has length dimensions minus
two.  The conformal factor $A(\vp)$ is
dimensionless.  In the case of a single scalar ($N=1$), the
target-space metric $\gamma_{AB}(\vp)$ reduces to a scalar function
$\gamma(\vp)$, and the choice $\gamma(\vp)=1$ can be made without loss
of generality.

The field equations of the theory, obtained by varying the action
(\ref{eq:action_ef}) with respect to $g^{\mu\nu}$ and $\vp$, take the form
\begin{align}
R_{\mu \nu} &= 2 \gamma_{AB}(\vp) \nabla_{\mu} \vp^{A} \nabla_{\nu} \vp^{B}
+ 2 V(\vp) g_{\mu \nu} + 8 \pi G_{\star} \left( T_{\mu \nu} - \frac{1}{2}Tg_{\mu \nu} \right) \,,
\label{eq:fieldeq_ef_tens}
\\
\Box \vp^{A} &=
- \gamma^{A}_{\phantom{A}BC}(\vp) g^{\mu \nu} \nabla_{\mu} \vp^{B} \nabla_{\nu} \vp^{C}
 +\gamma^{AB}(\vp)\frac{\partial V(\vp)}{\partial \vp^{B}}
 - 4 \pi G_{\star} \gamma^{AB}(\vp)\frac{\partial \log A(\vp)}{\partial \vp^{B}} T \,. \nonumber \\
\label{eq:fieldeq_ef_sc}
\end{align}
Here $\nabla_{\mu}$ is the covariant derivative associated with $g_{\mu \nu}$, and
$\Box \equiv \nabla^{\mu}\nabla_{\mu}$ is the corresponding d'Alembertian operator.
The Ricci tensor built out of the metric $g_{\mu \nu}$ is denoted
as $R_{\mu \nu}$.
The energy-momentum tensor $T_{\mu\nu}$ of the non-gravitational fields
is defined by
\begin{equation}
\label{eq:T}
T_{\mu \nu} \equiv -\frac{2}{\sqrt{-g}}\frac{\delta S_{\rm m} [A^{2}(\vp)g_{\rho \sigma} ; \Psi] }{\delta g^{\mu \nu}} \,,
\end{equation}
and its trace is given by $T \equiv g^{\mu \nu}T_{\mu \nu}$.

The energy conservation equation reads
\begin{equation}
\nabla^{\mu}\left[T_{\mu \nu} + T_{\mu \nu}^{(\vp)}\right] = 0 \,,
\end{equation}
or, more explicitly,
\begin{equation}
\nabla^{\mu} T_{\mu \nu} =\frac{\partial \log A(\vp)}{\partial \vp^{A}}\,  T \, \nabla_{\nu} \vp^{A} \,.
\end{equation}
Here
\begin{align}
\label{eq:Tphi}
T_{\mu \nu}^{(\vp)} &\equiv -\frac{2}{\sqrt{-g}}\frac{\delta S_{\vp} [g_{\rho \sigma} ; \varphi] }{\delta g^{\mu \nu}}
\nonumber
\\
&=
\frac{1}{4 \pi G_{\star}} \left[
\gamma_{AB}(\vp) \left(
\nabla_{\mu} \vp^{A} \nabla_{\nu}\vp^{B}
- \frac{1}{2} g_{\mu \nu} g^{\rho \sigma}
\nabla_{\rho} \vp^{A} \nabla_{\sigma} \vp^{B}
\right)
-V(\vp)g_{\mu \nu}
\right]
\end{align}
is an effective energy-momentum tensor for the scalar fields, where $S_{\vp}$ denotes
the scalar kinetic and potential contributions to the action (\ref{eq:action_ef}).
One may build an energy-momentum tensor which is conserved with
respect to the Levi-Civita connection of the Jordan-frame metric, and
whose components are directly related to physically observable
quantities as
\begin{equation}
  \tilde{T}_{\mu \nu} \equiv -\frac{2}{\sqrt{-\tilde{g}}}\frac{\delta S_{m} [\tilde{g}_{\rho \sigma} ;
      \Psi ] }{\delta \tilde{g}^{\mu \nu}} =
A^{-2}(\vp) T_{\mu \nu} \,.
\end{equation}
%
\subsection{Complexification}
\label{sec:complexification}
If the target space has a Hermitian structure, then it is useful to write the action in terms of holomorphic coordinates and
complexified tensors:
\begin{align}
S &= \frac{1}{4 \pi G_{\star}} \int d^{4}x \sqrt{-g} \left[ \frac{R}{4}
   -
 g^{\mu \nu} \gamma_{\bar{a}b}(\varphi,\bar{\varphi}) \nabla_{\mu}\bar{\varphi}^{a} \, \nabla_{\nu}\varphi^{b}
- V(\varphi,\bar{\varphi}) \right] \nonumber \\
 &+ S_{\rm m}[A^{2}(\varphi,\bar{\varphi})g_{\mu \nu} ; \Psi] \,.
\label{eq:action_ef_complex}
\end{align}
The complexified field equations are:
\begin{align}
R_{\mu \nu}  &= 4 \gamma_{\bar{a}b}(\varphi,\bar{\varphi}) \nabla_{(\mu} \bar{\varphi}^{a} \nabla_{\nu )} \varphi^{b}
+ 2 V(\varphi,\bar{\varphi}) g_{\mu \nu} + 8 \pi G_{\star} \left( T_{\mu \nu} - \frac{1}{2}Tg_{\mu \nu} \right) \,,
\label{eq:fieldeq_ef_tens_complex}
\\
\Box \varphi^{a}
&=
 -\gamma^{a}_{\phantom{a}bc}(\varphi,\bar{\varphi})g^{\mu \nu}\nabla_{\mu}\varphi^{b}\nabla_{\nu}\varphi^{c}
 - 2 \gamma^{a}_{\phantom{a}\bar{b}c}(\varphi,\bar{\varphi})
g^{\mu \nu}\nabla_{\mu}\bar{\varphi}^{b}\nabla_{\nu}\varphi^{c}
\nonumber
\\
&+\gamma^{a \bar{b}} (\varphi,\bar{\varphi}) \frac{\partial V(\varphi,\bar{\varphi})}{\partial \bar{\varphi}^{b}}
 - 4 \pi G_{\star} \gamma^{a \bar{b}}(\varphi,\bar{\varphi})\frac{\partial
   \log A(\varphi,\bar{\varphi})}{\partial \bar{\varphi}^{b}} T \,.
\label{eq:fieldeq_ef_sc_complex}
\end{align}
Note that for K\"{a}hler manifolds (and in particular one-complex-dimensional manifolds)
\begin{equation}
\gamma^{a}_{\phantom{a}\bar{b}c}(\varphi,\bar{\varphi})=0 \,, \qquad
\gamma^{a}_{\phantom{a}bc}(\varphi,\bar{\varphi}) = \gamma^{a\bar{d}}(\varphi,\bar{\varphi})
\frac{\partial \gamma_{c\bar{d}}(\varphi,\bar{\varphi})}{\partial \varphi^{b}} \,,
\end{equation}
so in this particular case the scalar field equations would simplify considerably.

\subsection{A two-real-scalar model with maximally symmetric target space}
\label{sec:theory-simple}
The simplest extension of a ST theory with a single real scalar field
is the case of two real scalar fields. We will mostly focus on this
model to illustrate the basic features of the new phenomenology
arising in TMS theories relative to the case of a single real scalar.
If the target space is assumed to be maximally symmetric, then there
are three possibilities for its geometry: flat, spherical, or
hyperbolic.  In the flat case, the target space may be trivially
identified with the complex plane $\mathbb{C}$.  In the spherical
case, the target space may be conformally mapped to the
one-point-compactification $\hat{\mathbb{C}}$ of the complex plane
$\mathbb{C}$ by means of stereographic projection.  In the case of a
hyperboloid of two sheets, the target space may be conformally mapped
to $\hat{\mathbb{C}} \backslash S^{1}$, also by means of stereographic
projection (we shall neglect the case of a hyperboloid of one
sheet); see \ref{app:targetspaces} for details.
Using the complex formulation discussed in Section~\ref{sec:complexification}, we work
with a single complex scalar rather than two real scalars, for which the
action~\eqref{eq:action_ef_complex} reduces to 
\begin{align}
S = \frac{1}{4 \pi G_{\star}} \int d^{4}x \sqrt{-g} \left[ \frac{R}{4}
-g^{\mu \nu} \gamma(\varphi,\bar{\varphi}) \nabla_{\mu}\bar{\varphi}
\, \nabla_{\nu}\varphi - V(\varphi,\bar{\varphi}) \right]
+ S_{\rm m}[A^{2}(\varphi,\bar{\varphi})g_{\mu \nu} ; \Psi] \,, \nonumber \\
\label{eq:action_ef_complex_single}
\end{align}
%
and the field equations are
\begin{align}
\label{eq:fieldeq_ef_tens_complex2}
R_{\mu \nu}  &= 4 \gamma(\varphi,\bar{\varphi}) \nabla_{(\mu} \bar{\varphi} \nabla_{\nu )} \varphi
+ 2 V(\varphi,\bar{\varphi}) g_{\mu \nu} + 8 \pi G_{\star} \left( T_{\mu \nu} - \frac{1}{2}Tg_{\mu \nu} \right) \,,
\\
\label{eq:fieldeq_ef_sc_complex2}
\Box \varphi
&=
 - \frac{\partial \log \gamma(\varphi,\bar{\varphi})}{\partial \varphi}g^{\mu \nu}\nabla_{\mu}\varphi\nabla_{\nu}\varphi
 +\gamma^{-1}(\varphi,\bar{\varphi})\frac{\partial V(\varphi,\bar{\varphi}) }{\partial \bar{\varphi}}
\nonumber
\\
&
 - 4 \pi G_{\star} \gamma^{-1}(\varphi,\bar{\varphi})\frac{\partial \log A(\varphi,\bar{\varphi})}{\partial \bar{\varphi}} T \,.
\end{align}
Hereafter we assume that the potential vanishes,
i.e. $V(\varphi,\bar{\varphi})=0$, and that the target space is
maximally symmetric. Therefore, upon stereographic projection and
field redefinition (see~\ref{app:targetspaces}),
the target-space metric can be written as
\begin{equation}
  \gamma(\varphi,\bar{\varphi}) = \frac{1}{2} \left( 1 + \frac{\bar{\varphi}\varphi}{4\gothr^2}\right)^{-2} \,,
  \label{targmetr}
\end{equation}
where $\gothr$ is the radius of curvature of the target-space
geometry: for a spherical geometry we have $\gothr^2>0$, for a
hyperbolic geometry $\gothr^2<0$, and in the limit
$\gothr \rightarrow \infty$ the geometry is flat.

With the above choices, the field equations become
\begin{align}
R_{\mu\nu} &= 2\left( 1 + \frac{\cvp\vp}{4\gothr^2} \right)^{-2}\partial_{(\mu}\cvp \partial_{\nu)}\vp +
8\pi G_{\star}\left( T_{\mu\nu} - \frac{1}{2}Tg_{\mu\nu} \right)\,,
\label{fieldeqR}\\
\Box\vp &=  \left( \frac{2\cvp}{\cvp\vp+4\gothr^2}\right)g^{\mu\nu} \partial_{\mu}\vp\partial_{\nu}\vp  -
  4\pi G_{\star}\left( 1+ \frac{\cvp\vp}{4\gothr^2}\right)\bar{\kappa}(\varphi,\bar{\varphi})\, T\,,
\label{fieldeqph}
\end{align}
where we introduced
\begin{equation}
\kappa(\varphi,\bar{\varphi}) \equiv 2 \left( 1 + \frac{\bar{\varphi} \varphi}{4 \radcurv^{2}}\right)
\frac{\partial \log A(\varphi,\bar{\varphi})}{\partial \varphi},
\label{eq:kappa_def}
\end{equation}
the so-called scalar-matter coupling function.

The function $A(\varphi,\bar{\varphi})$, whose derivative enters into
the field equations, determines the scalar-matter coupling through
Eq. (\ref{eq:kappa_def}).
Without loss of generality we assume that far away from the source the
field vanishes, i.e. that the asymptotic value of the scalar field is
$\varphi_\infty=0$.
We then expand the function $\log A$ in a series about $\varphi=0$:
%
\begin{equation}
\log A(\varphi, \bar{\varphi}) =\alpha^\ast\varphi+\bar{\alpha}^\ast\bar{\varphi}+
\frac{1}{2}\beta_0 \varphi \bar{\varphi} +
\frac{1}{4} \beta^\ast_1 \varphi^2  +
\frac{1}{4}\bar{\beta}^\ast_1 \bar{\varphi}^2 +\dots\,,
\label{eq:A_expansion}
\end{equation}
where $\beta_0$ is real, while $\alpha^\ast$ and $\beta^\ast_1$ are in
general complex numbers\footnote{At the onset of spontaneous
  scalarization $|\varphi|\ll1$, and we can always expand the
  conformal factor as in Eq.~\eqref{eq:A_expansion}. For scalarized
  solutions the field amplitude may be large, the higher-order terms
  in the expansion may not be negligible, and the
  expansion~\eqref{eq:A_expansion} should be considered as an ansatz
  for the conformal factor. For a general functional form of the
  conformal factor, the series expansion used here (and in
  Ref.~\cite{Damour:1993hw}) can only provide a qualitative
  description of the scalarized solution.}.
Although the five real parameters
${\rm Re}[\alpha^\ast], {\rm Im}[\alpha^\ast], \beta_{0}, {\rm
  Re}[\beta_{1}^\ast], {\rm Im}[\beta_{1}^\ast]$
are defined in terms of a specific target-space coordinate system, the
four real quantities
$(|\alpha^\ast|, \beta_0, |\beta_1^\ast|, \arg \alpha^\ast -
\frac{1}{2} \arg \beta_{1}^\ast)$
may be expressed solely in terms of target-space scalar quantities,
and thus have an invariant geometric meaning\footnote{The eigenvalues
  of the quadratic form in (\ref{eq:A_expansion}), given by
  $\beta_0 \pm |\beta_{1}^\ast|$, are target-space scalars. The phase
  difference $\arg \alpha^\ast - \frac{1}{2} \arg \beta_{1}^\ast$
  arises when this quadratic form is contracted with $\alpha^\ast$,
  see Eq.~(\ref{eq:argalpha}).}. The remaining real parameter is an
unmeasurable overall complex phase.
%

To make this explicit, redefine $\beta^\ast_1 \equiv \beta_1e^{\ii \theta}$, where
$\theta$ is chosen such that $\beta_1$ is real. 
Then, after defining $\alpha^\ast \equiv \alpha e^{\ii \theta/2}$ and
a new field $\psi \equiv \varphi e^{\ii \theta/2}$, the field
equations become
%
\begin{align}
R_{\mu\nu} &= 2\left( 1 + \frac{\bar{\psi}\psi}{4\gothr^2} \right)^{-2}\partial_{(\mu}\bar{\psi} \partial_{\nu)}\psi +
8\pi G_{\star}\left( T_{\mu\nu} - \frac{1}{2}Tg_{\mu\nu} \right) \,,
\label{Einstein_psi} \\
\Box\psi &= \left( \frac{2\bar{\psi}}{\bar{\psi}\psi+4\gothr^2}\right)g^{\mu\nu} \partial_{\mu}\psi\partial_{\nu}\psi
- 4\pi G_{\star}\left( 1+ \frac{\bar{\psi}\psi}{4\gothr^2}\right)
\bar{\kappa}(\psi,\bar{\psi})\,T\,,
\label{eq:psi}
\end{align}
%
where the function $\kappa$ is defined in Eq.~\eqref{eq:kappa_def} and
\begin{equation}
\log A(\psi, \bar{\psi}) =
\alpha\psi+\bar{\alpha}\bar{\psi}+
\frac{1}{2}\beta_0 \psi \bar{\psi} +
\frac{1}{4}\beta_1 \psi^2  +
\frac{1}{4}\beta_1 \bar{\psi}^2
+\dots \,.
\label{eq:logA}
\end{equation}
Therefore, any solution of the original theory (formulated with respect to $\varphi$ and complex coupling coefficients
$\alpha^\ast$ and $\beta^\ast_1$) can be obtained from a theory where we consider the field $\psi$, a
real-valued $\beta_1$ and a generically complex $\alpha$. 
The solution for the theory corresponding to the conformal
factor~\eqref{eq:A_expansion} is then given by a simple rotation,
$\varphi=\psi \exp\left(-\ii \theta\right /2)$.

The model just described represents the simplest, yet quite
comprehensive, generalization of the model of single ST theory
investigated originally in Ref.~\cite{Damour:1993hw}.

Note that the quantity
$|\alpha|^2 \equiv \alpha\bar{\alpha}\equiv
\rm{Re}[\alpha]^2+\rm{Im}[\alpha]^2$
is strongly constrained by observations (cf.~\ref{app:constraints}),
similarly to the single-scalar case. When $\alpha=0$, the conformal
coupling reduces to
\begin{equation}
\log A(\psi, \bar{\psi}) =
\frac{1}{2}\beta_0 \psi \bar{\psi} +
\frac{1}{4}\beta_1 \psi^2  +
\frac{1}{4}\beta_1 \bar{\psi}^2 \,,
\label{cf0}
\end{equation}
where we neglected higher-order terms in the scalar field.
However, in TMS theories $\alpha$ is a complex quantity and its
argument, $\arg\alpha$, is completely unconstrained in the weak-field
regime.
In Section~\ref{sec:full_theory} we will show that compact
stars in theories with $\alpha=0$ and $\alpha\neq0$ are rather
different.

The field equations can be also written in terms of two real
scalars. For this purpose, let us split the field $\psi$ into real and
imaginary parts: $\psi\equiv {\rm Re}[\psi] + \ii \,{\rm
  Im}[\psi]$.
Then the conformal factor~\eqref{cf0}, again in the
$\alpha=\bar\alpha=0$ case, reads:

\begin{equation}
\log A(\psi, \bar{\psi}) = \frac{1}{2} \left[
(\beta_0+\beta_1) {\rm Re}[\psi]^2 +
(\beta_0-\beta_1) {\rm Im}[\psi]^2  \right]\,.   \label{cf:realsplit}
\end{equation}
The structure of this TMS theory is ultimately determined by
three real parameters: $\beta_0+\beta_1$, $\beta_0-\beta_1$ and the
target-space curvature defined by $\gothr^2$. When $\alpha\neq0$, two
further parameters ($|\alpha|$ and $\arg\alpha$) are necessary to
define the theory.

\section{Stellar structure in tensor-multi-scalar theories}
\label{sec:stellarstructure}

In this section we consider the structure of relativistic stars in the
context of the TMS theory introduced in
Section~\ref{sec:theory-simple}. We first derive the equations of
structure for a slowly rotating star in the Hartle-Thorne
formalism~\cite{Hartle:1967he,Hartle:1968ht} (Section~\ref{sec:ss_eqs}),
then we integrate these equations and discuss some properties of
scalarized solutions in increasingly complex scenarios
(Section~\ref{sec:ss_num}).

\subsection{Equations of hydrostatic equilibrium}
\label{sec:ss_eqs}

We describe a stationary, axisymmetric star, composed by a perfect fluid, slowly
rotating with angular velocity $\Omega$, using coordinates
$x^{\mu} = (t, r, \theta, \phi)$ and the line element
\begin{align}
g_{\mu \nu}dx^{\mu} dx^{\nu} &=
-e^{\nu(r)}dt^{2} + \frac{dr^{2}}{1-2 \mu(r)}
 + r^{2}(d\theta^{2} + \sin^{2}\theta d\phi^{2})
\nonumber
\\
&
+ 2 \left[\omega (r) - \Omega\right] r^{2} \sin^2 \theta dt d\phi.
\end{align}
where we neglect terms of order $\sim\Omega^2$ and higher in the
metric and in the hydrodynamical quantities.  The variable $\mu(r)$ is
related to the more familiar mass function $m(r)$ by $\mu = m/r$.  The
energy-momentum tensor of the perfect fluid takes the usual form
\begin{equation}
T^{\mu \nu} = A^{4}(\psi,\bar{\psi})
\left[
(\rho + P)u^{\mu} u^{\nu} + P g^{\mu \nu}
 \right] \,,
\end{equation}
where $\rho$, $P$, and
$\tilde{u}^{\mu} = A^{-1}(\psi,\bar{\psi})u^{\mu}$ are the mass-energy
density, pressure, and four-velocity of the fluid, respectively, and
\begin{equation}
u^{\mu} = e^{-\nu/2} ( 1,\, 0, \,0, \,\Omega) \,.
\end{equation}
With these choices, the field equations
(\ref{eq:fieldeq_ef_tens})--(\ref{eq:fieldeq_ef_sc}) reduce to a system
of coupled ordinary differential equations, namely
\begin{align}
(r \mu)' &=
\frac{1}{2}(1-2 \mu)r^{2}
\left(1 + \frac{\bar{\psi}\psi}{4\radcurv^{2}}\right)^{-2} \bar{\psi}' \psi'
+ {4 \pi \bareG} A^{4} (\psi,\bar{\psi}) r^{2} \massdens \,,
\\
P' &= - (\massdens c^{2} + P)
\left\{ \frac{\nu'}{2} + \frac{1}{2} \left( 1 + \frac{\bar{\psi} \psi}{4\radcurv^{2}}\right)^{-1}
\left[\kappa(\psi,\bar{\psi}) \psi' + \bar{\kappa}(\psi,\bar{\psi}) \bar{\psi}'\right] \right\} \,,
\\
\nu' &= \frac{2\mu}{r(1-2\mu)}
+ r \left( 1 + \frac{\bar{\psi} \psi}{4\radcurv^{2}} \right)^{-2} \bar{\psi}' \psi'
+ \frac{8 \pi \bareG A^{4}(\psi,\bar{\psi})r P}{1-2\mu} \,,
\\
\psi'' &=
\frac{2\bar{\psi} \psi'^{2}}{\bar{\psi}\psi + 4\radcurv^{2}}
-  \frac{2(1-\mu)}{r(1-2\mu)} \psi'
\nonumber
\\
&
+\frac{4 \pi \bareG A^{4}(\psi,\bar{\psi})}{1-2\mu} \left[
(\massdens -3P) \left( 1 + \frac{\bar{\psi}\psi}{4\radcurv^{2}} \right) \bar{\kappa}(\psi,\bar{\psi})
+r \psi' (\massdens - P)
\right] \,,
\\
\omega'' &=
\left[ r \left( 1 + \frac{\bar{\psi} \psi}{4\radcurv^{2}} \right)^{-2} \bar{\psi}' \psi'
- \frac{4}{r} \right] \omega'
+\frac{4 \pi \bareG A^{4}(\psi,\bar{\psi})r(\rho c^{2} + P)}{1-2\mu}
\left( \omega' + \frac{4}{r}\omega\right) \,, \nonumber \\
\end{align}
where primes denote derivatives with respect to the radial coordinate
$x^{1}=r$.  As usual, the system is closed by specifying a barotropic
equation of state $P=P(\rho)$.

For the purpose of numerical integration, it is useful to work out series expansions
of the functions $\mu(r)$, $\nu(r)$, $P(r)$, $\psi(r)$, and $\omega(r)$ about $r=0$:
\begin{align}
\mu(r) &= \frac{4 \pi \bareG}{3}A^{4}_{0}\rho_{0}r^{2}
+ \mathcal{O}(r^{4}) \,,
\\
\nu(r) &= \nu_{0} + \frac{4 \pi \bareG}{3}A^{4}_{0} (\rho_{0} +3P_{0})r^{2}
+ \mathcal{O}(r^{4}) \,,
\\
P(r) &= P_{0} - \frac{2 \pi \bareG}{3}A^{4}_{0}(\rho_{0} + P_{0})
\left[\rho_{0} + 3P_{0} + \bar{\kappa}_{0}\kappa_{0}\left(\rho_{0}-3P_{0}\right)\right]r^{2}
+ \mathcal{O}(r^{4}) \,, \\
\psi(r) &= \psi_{0} + \frac{2 \pi \bareG}{3} A^{4}_{0}\bar{\kappa}_{0}
(\rho_{0} - 3P_{0})\left(1 + \frac{\bar{\psi}_{0}\psi_{0}}{4\radcurv^{2}}\right)r^{2}
+ \mathcal{O}(r^{4}) \,,
\label{eq:scalar_field_r0}
\\
\omega(r) &= \omega_{0} + \frac{8 \pi \bareG}{5} \omega_{0} A^{4}_{0}
(\rho_{0} + P_{0})r^{2}
+ \mathcal{O}(r^{4}) \,.
\end{align}
Here the subscript $0$ denotes evaluation at the stellar center $r=0$.

When $\psi$ and $\bar\psi$ are constant and $A(\psi,\bar\psi)=1$ the
field equations reduce to the standard Tolman-Oppenheimer-Volkoff
equations in GR. Indeed, GR solutions are part of the solution
spectrum of TMS theories with $\alpha=0$ in Eq.~\eqref{eq:logA}, as in
the usual single-scalar case~\cite{Damour:1992we,Damour:1993hw}. On
the other hand, when $\alpha\neq0$ the scalar field is forced to have
a non-trivial profile in the presence of matter ($T\neq0$).

Furthermore, even when $\alpha=0$, other solutions characterized by a
non-trivial profile for the scalar fields can co-exist with the GR
solutions; in~\ref{app:linear} we give a simple interpretation of
these ``spontaneously scalarized'' solutions in terms of a tachyonic
instability of relativistic GR solutions. Besides their gravitational
mass $M$ and radius $R$, scalarized neutron stars are characterized by
their scalar charge $Q$, which is generally a complex number for the
complex scalar field $\psi$ discussed here\footnote{Note that $Q$ is
  defined in terms of a specific target-space coordinate
  system. Observable quantities must be invariant under changes of
  coordinates, and can thus only depend on target-space scalars such
  as $|Q|^2$, or $|Q_{\ssA}-Q_{\ssB}|^2$ (for a binary system with
  bodies A and B), or more complicated contractions.}.

The gravitational mass $M$ and the scalar charge $Q$ of stellar models
in TMS theories can be computed by integration in the vacuum exterior region,
where $P = \rho = 0$.
We integrate the structure equations outwards from $r = 0$ up to a
point $r = R$ such that $P(R) = 0$, which determines the
stellar surface in the Einstein frame. The areal radius $\tilde{R}$ in the Jordan frame can be obtained rescaling $R$ by the conformal
factor $A(\psi, \bar{\psi})$, which depends on the value of the
scalar field $\psi$ and its complex conjugate $\bar{\psi}$ at $r = R$.
The values of the mass function $m$, of the scalar field $\psi$ and
of its derivative $\psi^{\prime}$ at $r = R$ are used as initial
conditions to integrate the structure equations in the exterior region. In
practice, the integration is terminated at some finite but large grid
point where $r = R_{\infty}$. From the values of $m$, $\psi$ and
$\psi^{\prime}$ at $R_{\infty}$ we can determine $M$ and $Q$ by
solving the system of equations

\begin{align}
m(r) =& M - \frac{|Q|^2}{2r} - \frac{M|Q|^2}{2r^2} + \mathcal{O}\left(\frac{1}{r^3}\right) \,,
\\
\psi(r) =& \psi_{\infty} + \frac{Q}{r} + \frac{QM}{r^2}  + \mathcal{O}\left(\frac{1}{r^3}\right) \,,
\\
\psi^{\prime}(r) =& -\frac{Q}{r^2} + {\cal O}\left( \frac{1}{r^3} \right)\,.
\end{align}
where $\psi_{\infty}$ is the constant background value of the scalar
field. Therefore we have three equations to solve for three unknowns:
$M$, $Q$ and $\psi_{\infty}$.
The physical solution of interest is
the one corresponding to the particular central value of the scalar
field $\psi_0$ (which can be found e.g. by a shooting method) such
that the background field vanishes, i.e. $\psi_{\infty} = 0$.

As an alternative integration technique, we have also implemented a
compactified coordinate grid in the vacuum exterior introducing a variable
$y\equiv 1/r$, which results in a regular set of differential
equations that is readily integrated to spatial infinity at $y=0$. The
gravitational mass $M$, scalar charge $Q$ and asymptotic scalar field
magnitude are then directly obtained from $m(y=0)$, $\psi(y=0)$ and
$\partial_y \psi (y=0)$, and numerical shooting provides a
fast-converging algorithm to enforce the boundary condition
$\lim_{y\rightarrow 0} \psi = 0$.  The two independent integrators
yield bulk stellar properties that agree within $\sim 1\%$ or better.

\subsection{Numerical integration and results}
\label{sec:ss_num}

In this section we discuss the result of the numerical integration of
the hydrostatic equilibrium equations in the TMS theory of
Section~\ref{sec:theory-simple}. Our main interest is to understand
scalarization in this model, and for simplicity in this paper we will
focus on static stars. We will use the polytropic equation of state
labeled ``EOS1'' in Ref.~\cite{Novak:1997hw}, for which the pressure $P$
and the energy density $\rho$ are given as functions of the baryonic
density $n_B$ by
\begin{eqnarray}
  P = K n_0 m_B \left( \frac{n_B}{n_0} \right)^{\gamma}\,,
    \qquad
  \rho = n_B \,m_B + K \frac{n_0 m_B}{\gamma-1}
        \left( \frac{n_B}{n_0} \right)^{\gamma}\,,
        \label{eq:EOS_rho}
\end{eqnarray}
where $n_0 = 0.1~{\rm fm}^{-3}$, $m_B=1.66 \times 10^{-27}~{\rm kg}$,
$K=0.0195$ and $\gamma=2.34$. Therefore, the function $\rho=\rho(P)$
can be constructed parametrically by varying $n_B$.

\subsubsection{The $O(2)$-symmetric theory}
\label{sec:u1}
In the absence of a scalar potential, the gravitational part of the
action \eqref{eq:action_ef} is invariant under the target-space
isometry group $G$. For our simple two-real-scalar model with
maximally symmetric target space, $G$ is the orthogonal group $O(3)$
in the case of spherical geometry, the indefinite orthogonal group
$O(2,1)$ in the case of hyperbolic geometry, and the inhomogeneous
orthogonal group $IO(2) = \mathbb{R}^{2} \rtimes O(2)$ in the case of
flat geometry.

When scalar-matter couplings are introduced, the action is no longer
invariant under all of $G$, but only under some subgroup $H < G$.  As
a first example, let us consider the particular case in which
$\beta_1=\alpha=0$. In this case the conformal factor
$A(\psi, \bar{\psi})$ given in Eq.~\eqref{cf0} reduces to
\begin{equation}
A(\psi, \bar{\psi}) = \exp\left(\frac{1}{2}\beta_0 \psi \bar{\psi}\right)\,,
\end{equation}
where again we have neglected higher-order terms in the scalar field.
This equation is obviously invariant under rotations in the complex plane
($\psi \to \psi\, e^{\ii \theta}$) and complex conjugation
($\psi \to \bar{\psi}$). Therefore, $H=O(2)$.  Note that the boundary
condition $\psi_{\infty}=0$ is $H$-invariant.  We refer to this
special case as the $O(2)$-symmetric TMS theory. In this theory, a GR
stellar configuration with $\psi \equiv 0$ is always a solution that
is $O(2)$-invariant.

\begin{figure}[thb!]
\centering
\includegraphics[width=\textwidth]{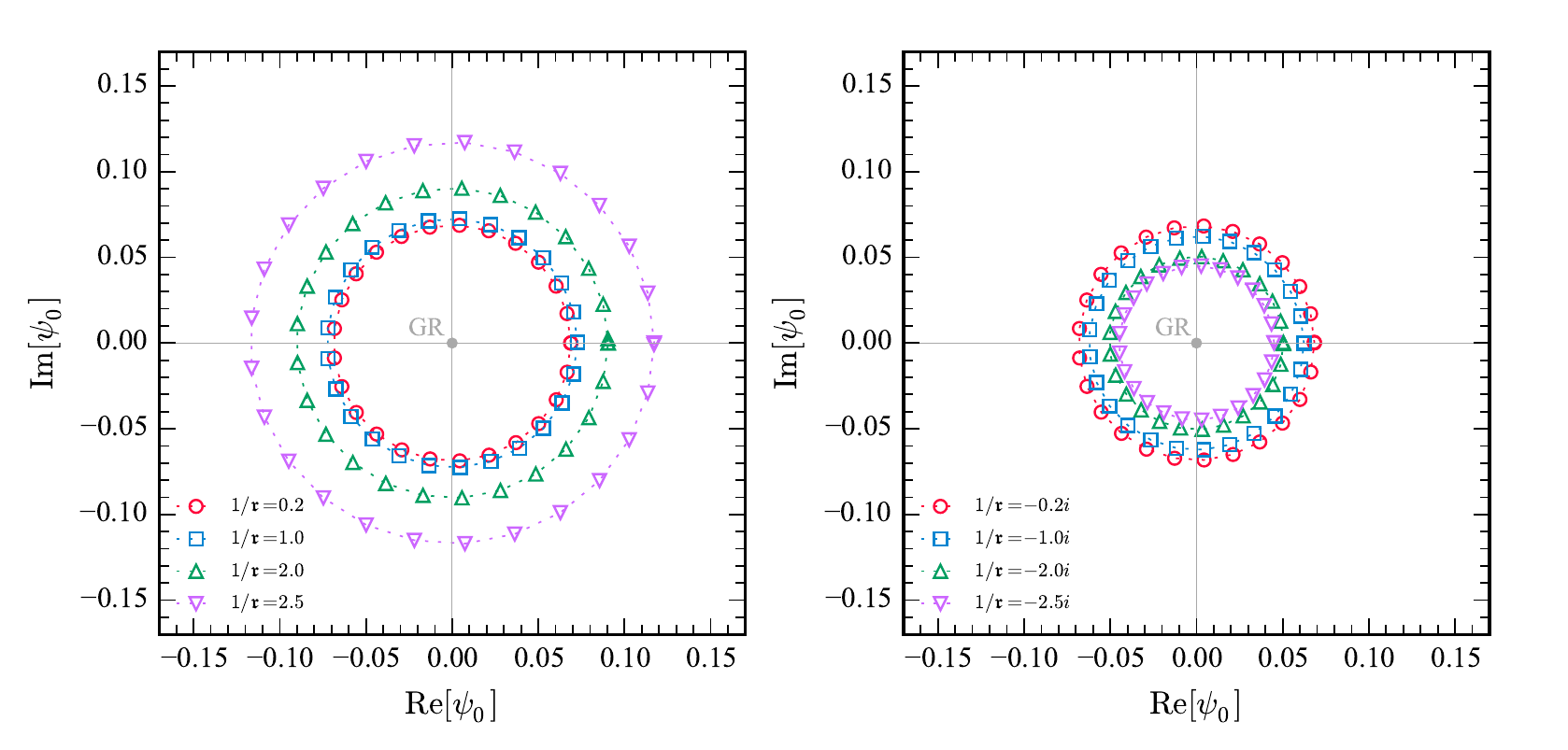}
\caption{{\it Spontaneous scalarization in a TMS theory with $O(2)$
    symmetry.} The value $\psi_0$ of the scalar field at the center of
  the star for scalarized solutions in the $O(2)$--symmetric theory
  with $\beta_0 = -5.0$ and central baryon density
  $n_{B} = 10.4 \, n_{\rm nuc}$, where the nuclear density is
  $n_{\rm nuc} = 10^{44} \, {\rm m}^{-3}$.  {\it Left panel:}
  spherical target space with $\radcurv^2 > 0$.  {\it Right panel:}
  hyperbolic target space with $\radcurv^2 < 0$. In both panels the
  origin corresponds to the neutron star solution in GR.  }
\label{fig:u1_sym}
\end{figure}

\begin{figure}[ht]
\includegraphics[width=\textwidth]{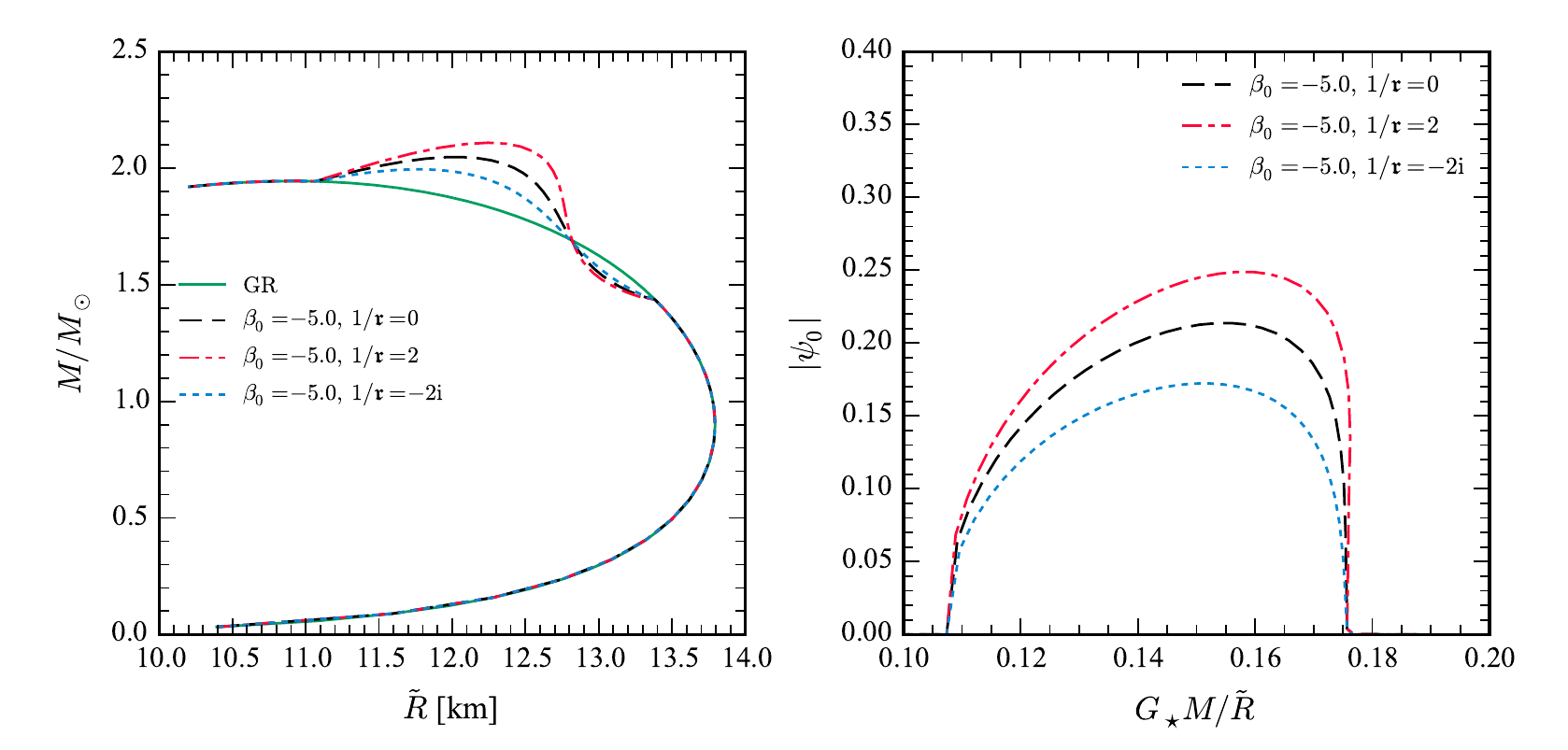}
\caption{{\it Stellar properties in the $O(2)$--symmetric theory.}  {\it
    Left panel}: The mass-radius relation for different values of
  $\radcurv$ and $\beta_0 = -5.0$. {\it Right panel}: Central value
  of the magnitude of the scalar field $\vert \psi_0\vert$ as a function
  of the stellar compactness $\bareG M/(\tilde{R} c^2)$.
  Here $\tilde{R}$ is the areal Jordan-frame radius of the star. The onset of
  scalarization does not  depend on the value of $\radcurv$.}
\label{fig:u_1_mass_radius}
\end{figure}

We now construct scalarized solutions, which spontaneously break the
$O(2)$ symmetry.  They depend on the two real parameters ($\beta_0$
and $\gothr^2$) of this theory, as well as the central baryon density
$n_{B}$.  The $O(2)$--symmetric character of the scalarized solution
space is exhibited in Fig.~\ref{fig:u1_sym}, where we show that, for
given values of $\gothr$ and $n_{B}$, there exists an infinite number
of scalarized solutions characterized by a different value of the
complex field $\psi_0$ at the center of the star.  The different
values of the scalar field are related by a phase rotation, and the
masses and radii of neutron star models along each of the circles
shown in Fig.~\ref{fig:u1_sym} are identical. The target-space
curvature $\radcurv$ has the effect of suppressing ($\radcurv^2 < 0$)
or increasing ($\radcurv^2 > 0$) the value of $|\psi_0|$, and
consequently of the scalar charge $Q$. Therefore a spherical target
space ($\radcurv^2 > 0$) produces stronger scalarization effects in
the mass-radius relations with respect to the case of a flat
target-space metric, as illustrated in
Fig.~\ref{fig:u_1_mass_radius}. On the other hand, a hyperbolic target
space ($\radcurv^2 < 0$) tends to reduce the effects of spontaneous
scalarization.
This can be intuitively, if not rigorously, understood by a glance at
Eqs.~\eqref{fieldeqph} and \eqref{eq:kappa_def}: the curvature term
plays the role of an ``effective (field-dependent) gravitational
constant'' which is either larger or smaller than the ``bare''
gravitational constant depending on whether $\radcurv^2>0$ or
$\radcurv^2<0$.
In both cases, as $\gothr\to\infty$ the solution reduces (modulo a
trivial phase rotation) to that of a ST theory with a single real
scalar field $\psi$ and scalar-matter coupling
$A(\psi) = \exp\left(\frac{1}{2}\beta_0 \psi^2\right)$.  We remark
that due to the $O(2)$ symmetry, all solutions of this theory are
equivalent to solutions with ${\rm Im}[\psi]=0$; as discussed in
Section~\ref{sec:full_theory} below, these are effectively -- modulo a
field redefinition -- solutions of a single-scalar theory.

\begin{figure}[ht]
\includegraphics[width=\textwidth]{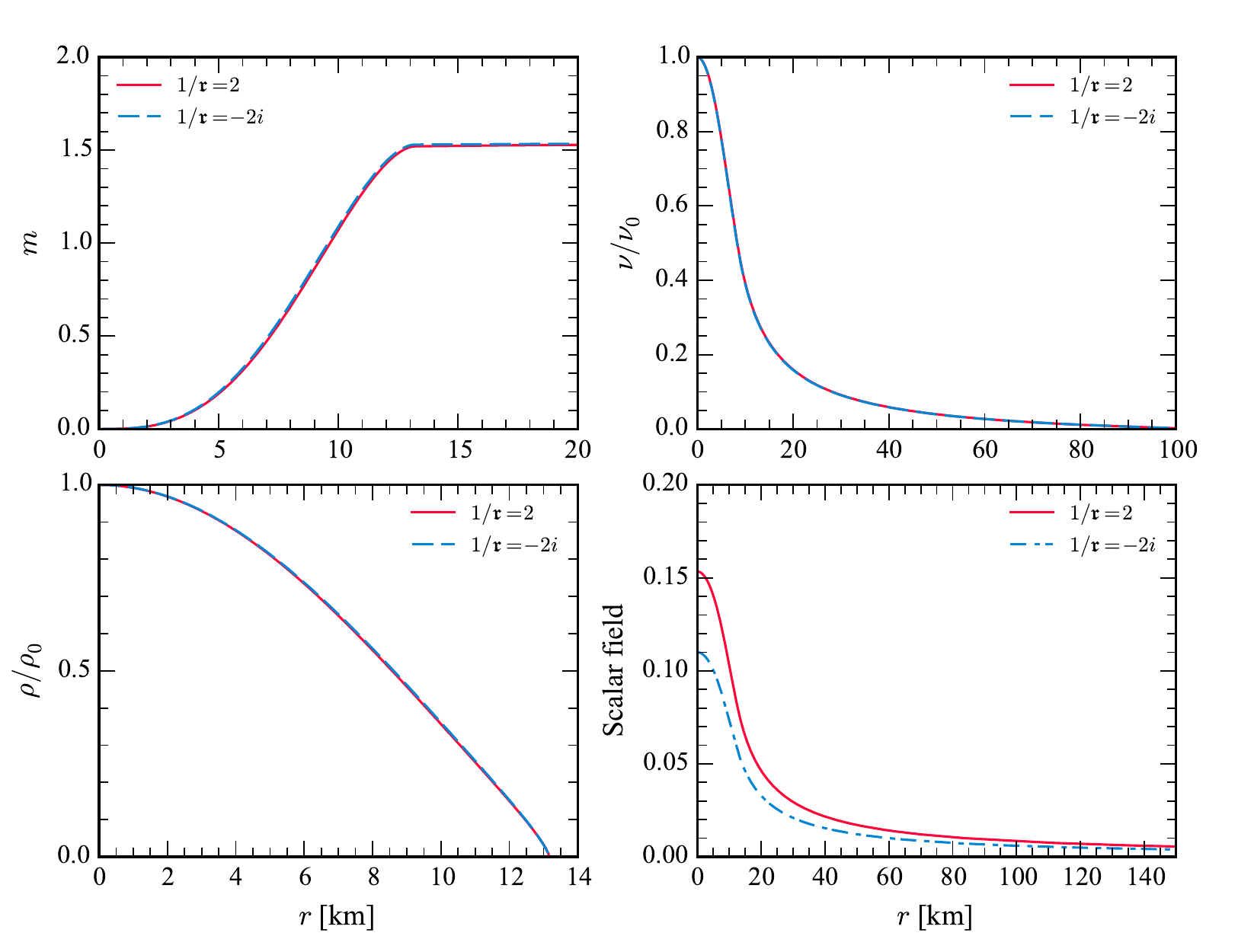}
\caption{
{\it Radial profiles.}
Different panels show the profiles of the mass function $m$, metric
potential $\nu$, total energy density $\rho$ and complex
scalar field $\psi$, in units of $c = G = M_{\odot} =1$.
The profiles correspond to a scalarized
star in the $O(2)$-symmetric theory with $\beta_0 = -5.0$,
$\beta_1 = \alpha = 0$,
and fixed baryon mass $M_B = 1.70$ $M_{\odot}$.
The target space curvature is either $\radcurv = 0.5$ (spherical)
or $\radcurv = 0.5 \ii$ (hyperboloidal).
In the spherical case, the scalarized solution has a gravitational mass
$M = 1.54$ $M_{\odot}$, Jordan-frame areal radius $\tilde{R} = 13.0$ km,
total scalar charge $Q = 0.553$ $M_{\odot}$ and central scalar magnitude $|\psi_0| = 0.154$. In the hyperbolic case, these
quantities are $M = 1.54$ $M_{\odot}$, $\tilde{R} = 13.0$ km,
$Q = 0.393$ $M_{\odot}$ and $|\psi_0| = 0.110$. For comparison, the GR
solution with the same baryonic mass has $M = 1.54$ $M_{\odot}$ and
$R = \tilde{R} = 13.2$ km.
}
\label{fig:u_1_profiles}
\end{figure}

Finally, in Fig.~\ref{fig:u_1_profiles} we illustrate the radial
profiles of the mass function $m$, metric potential $\nu$, mass-energy
density $\rho$ and scalar field $\psi$ for scalarized stellar models
with fixed baryonic mass $M_B = 1.70$ $M_{\odot}$ in theories with
$\beta_0 = -5.0$ and $\radcurv^2 = \pm 1 / 4$.

\subsubsection{The full TMS theory}
\label{sec:full_theory}
We now turn our attention to the existence of scalarized stellar
models in the theory defined by Eq.~\eqref{eq:logA}, which depends on
three real parameters ($\beta_0$, $\beta_1$ and $\gothr^2$) and the
complex constant $\alpha$.  When $\alpha = 0$ and $\beta_{1} \neq 0$, this theory is
invariant under the symmetry group $Z_{2} \times Z_{2}$ generated by
conjugation ($\psi \to \bar{\psi}$) and inversion
($\psi \to - \psi$).  Introduction of $\alpha \in \mathbb{R}$
partially breaks this symmetry down to $Z_{2}$, consisting of
conjugation only, whereas introduction of
$\alpha \in \mathbb{C} \backslash \mathbb{R}$ fully breaks this
symmetry.

An interesting question is whether there exists a region of the
parameter space of this theory in which \emph{both} fields
scalarize\footnote{This question is not invariant under field
  redefinitions. More precisely, we ask whether there exists a doubly
  scalarized solution which can not be described by an effective
  single-real-scalar theory.}.
We first searched for such ``biscalarized'' solutions in the
$Z_{2} \times Z_{2}$ theory with $\alpha = 0$, considering a wide range of the
($\beta_0,\beta_1,\gothr$) space, but we could not find any.
However, the situation is dramatically different when
$\alpha\neq0$. Crucially, $|\alpha|$ has to be small enough to satisfy
the observational bounds summarized in~\ref{app:constraints}, and in
particular Eq.~\eqref{boundalpha}, but $\arg{\alpha}$ is completely
unconstrained by weak-field observations.
Our numerical findings are in agreement with an approximate analytical
model which will be presented elsewhere~\cite{elsewhere}.
In the following we discuss the cases $\alpha=0$ and $\alpha\neq0$
separately.

\paragraph{Case $\alpha=0$: breaking the $O(2)$ symmetry down to $Z_{2} \times Z_{2}$}

\begin{figure}[ht]
\begin{center}
\includegraphics[width=0.6\textwidth]{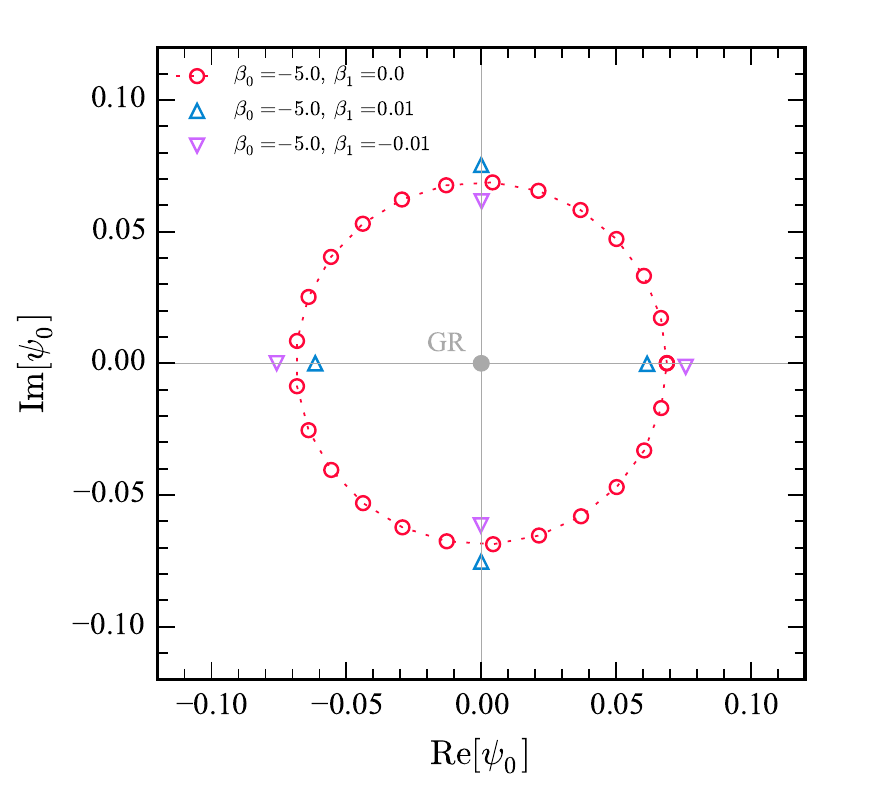}
\caption{{\it Symmetry breaking of the space of solutions.} When
  $\beta_1\neq0$, the $O(2)$-symmetric solution-space analyzed in the
  previous section (cf. Fig.~\ref{fig:u1_sym}) collapses down to a $(Z_{2} \times Z_{2})$-symmetric solution-space.
  This property of the theory
  is here illustrated for stellar models with the same equation of state and central
  energy density as in Fig.~\ref{fig:u1_sym}, $\beta_0 = -5.0$ and
  $\radcurv = 5.0$.  }
\label{fig:u_1_break}
\end{center}
\end{figure}

When $\alpha=0$ but $\beta_1\ne 0$, we found solutions where only
\emph{either the real or the imaginary part} of the scalar field has a
non-trivial profile. Therefore, in this case the circle shown in
Fig.~\ref{fig:u1_sym} for the $O(2)$-symmetric theory collapses down to
four discrete points on the real and imaginary axes
(cf. Fig.~\ref{fig:u_1_break}).

\begin{figure}
  \centering
  \includegraphics[width=0.49\textwidth]{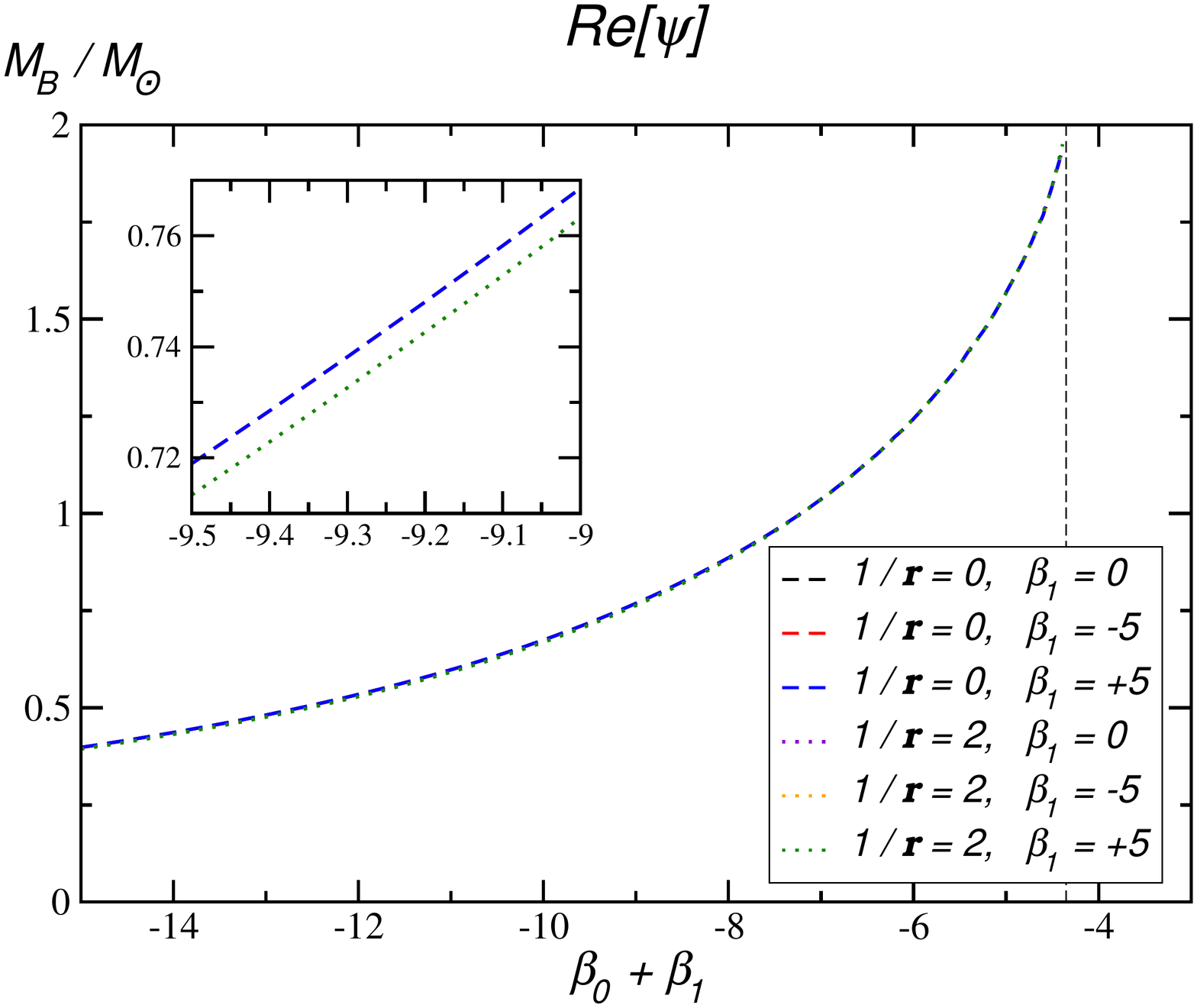}
  \includegraphics[width=0.49\textwidth]{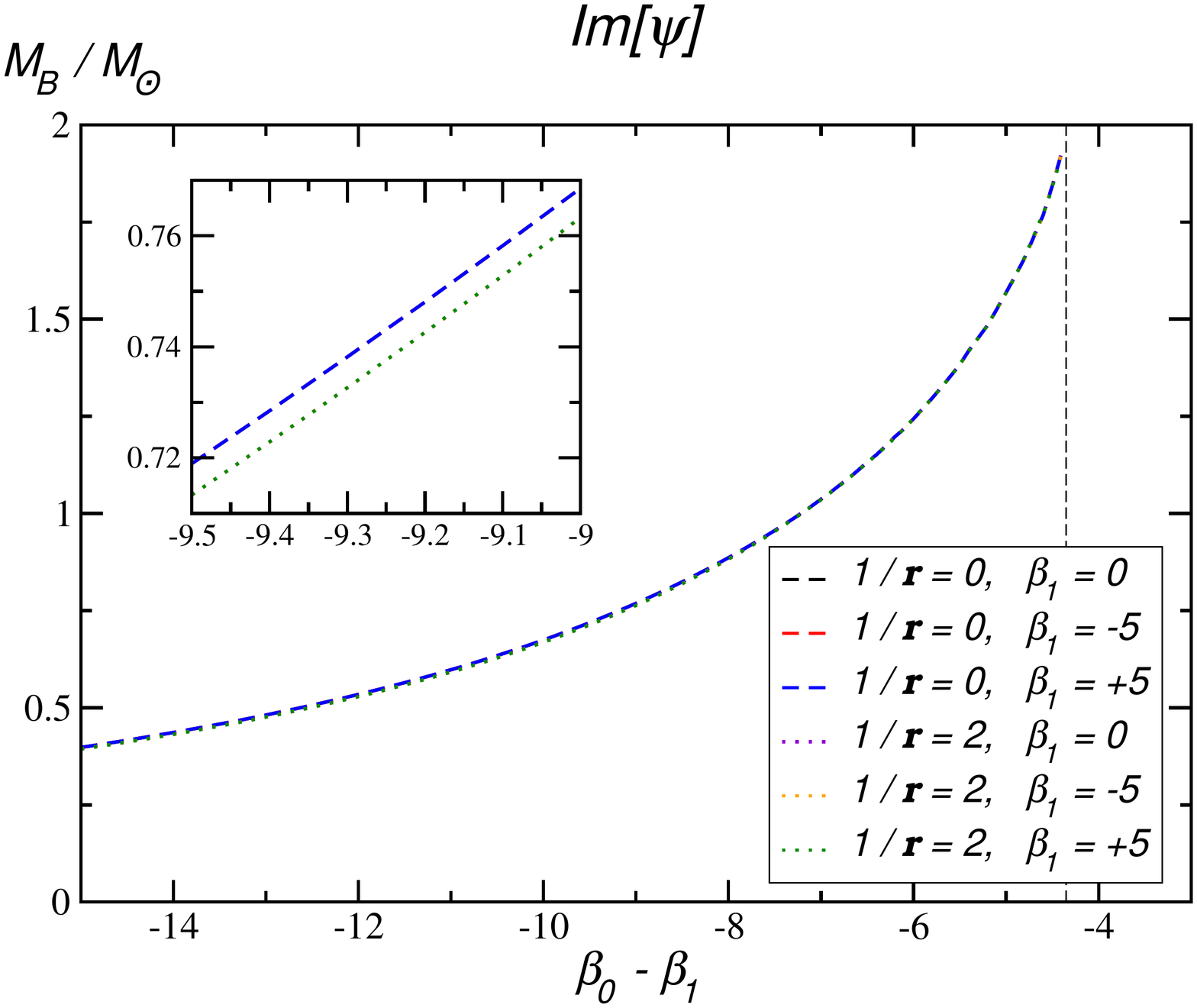}
  \caption{ {\it Minimal baryonic mass of scalarized models.}  The
    baryon mass of scalarized solutions at the onset of scalarization
    as a function of (i) $\beta_0 + \beta_1$ for models where
    ${\rm Re}[\psi]$ is non-zero (left panel) and (ii)
    $\beta_0 - \beta_1$ for models where ${\rm Im}[\psi]$ is non-zero
    (right panel).  Each panel contains 6 curves, corresponding to the
    three values of $\beta_1$ at fixed $\radcurv = \infty$ (dashed
    curves) and the same three values of $\beta_1$ at $\radcurv = 1/2$
    (dotted curves). The three dashed curves and the three dotted
    curves, respectively, are indistinguishable in the plot and the
    two families of dashed and dotted curves are only distinguishable
    in the inset, where we zoom into a smaller region. In both panels,
    the vertical long-dashed curve denotes the value
    $\beta_0 \pm \beta_1 = -4.35$ above which we no longer identify
    scalarized models, in agreement with
    Eq.~(\ref{doublescalarization}).  From Eq.~\eqref{cf:realsplit} it
    is clear that the natural parameters are $\beta_0+\beta_1$ and
    $\beta_0-\beta_1$ when the theory is written in terms of the real
    and imaginary part of $\psi$, respectively.}
  \label{fig:scalarization}
\end{figure}

In~\ref{app:linear} we perform a linear analysis of the field
equations, deriving the conditions for scalarization to occur.
From Eqs.~(\ref{eqlinZ}), (\ref{eqlinW}) and
(\ref{doublescalarization}) we expect that scalarized models exist if
$\beta_0 + \beta_1 \lesssim -4.35$ when ${\rm Re}[\psi]\neq 0$, or
$\beta_0 - \beta_1 \lesssim -4.35$ when instead
${\rm Im}[\psi]\neq 0$. We have checked this expectation by
calculating models for the parameter sets (i) $1/\radcurv = 0$,
$\beta_1 = 0$ and (ii) $1/\radcurv = 2$, $\beta_1 = 0$.
For each of these cases, we have varied the central density from
$10^{-5}~{\rm km}^{-2}$ to $0.0015~{\rm km}^{-2}$ in steps of
$10^{-5}~{\rm km}^{-2}$. We applied our shooting algorithm for a
scalar field amplitude $|\psi(r=0)| \in [0,1]$ in steps of $0.1$,
choosing discrete values of the complex phase
$\theta = 0,\,\pi/2,\,\pi,\,3\pi/2$, and varying $\beta_0 \in [-20,3]$
in steps of $0.01$.
For all values of the central density and $\beta_0$, the shooting
method identifies one GR solution model with vanishing scalar charge.
For sufficiently negative $\beta_0$, we additionally identify scalarized
models.  Among these models we then identify for a given value of
$\beta_0$ the scalarized model with the lowest baryon mass, and thus
generate a scalarization plot analogous to Fig.~2
in~\cite{Damour:1996ke} for ST theory with a single scalar field. The
result is shown in Fig.~\ref{fig:scalarization}. 
The small difference between the curves for different curvature radius
$\radcurv$ likely arises from the small but finite amplitude of the
scalar field appearing in the lowest-mass scalarized binaries, which
is a byproduct of finite discretization in the mass parameter
space. In the continuum limit of infinitesimal amplitudes of the
scalar field in scalarized models, we expect this difference to
disappear completely and the dotted and dashed curves to overlap. This
is indeed supported by an analytic calculation.\footnote{This
  calculation uses Riemann-normal coordinates at $\varphi_{\infty}$ in
  target space, and finds that target-space-curvature terms appear in
  the field equations at third order in the scalar-amplitude
  expansion. Details will be published elsewhere~\cite{elsewhere}.}
These results confirm the prediction of
Eq.~(\ref{doublescalarization}) and agree (qualitatively and
quantitatively) with the single-scalar case shown in Fig.~2
of~\cite{Damour:1996ke}.
Indeed, in this case the analogy with the single-scalar case can be
made more formal. Let us consider without loss of generality the
subspace of the solution space in which the scalar field is real,
i.e. $Z={\rm Re}[\psi]\neq 0$, $W={\rm Im}[\psi]=0$.
The kinetic term can be put in the canonical form by a scalar field
redefinition, i.e.
\be
K=-\frac{1}{2}\left(1+\frac{Z^2}{4\gothr^2}\right)^{-2}\partial_\mu Z\partial^\mu Z=
-\frac{1}{2}\partial_\mu \tilde Z\partial^\mu \tilde Z\,,
\label{eq:Kscal}
\ee
where the two fields are related by
$Z=2\gothr\tan\left(\frac{\tilde Z}{2\gothr}\right)$,
and $-\pi\gothr<\tilde Z<\pi\gothr$. For $|Z|\ll\gothr$ we have
\begin{align}
\tilde Z&=Z-\frac{Z^3}{12\gothr^2}+O(Z^5)\,,\qquad Z=\tilde Z+\frac{\tilde Z^3}{12\gothr^2}+O(\tilde Z^5)\,.
\label{eq:Zscal}
\end{align}
Replacing this Taylor expansion in the conformal factor~\eqref{cf0} we
see that the parameters $\beta_0,\beta_1$ remain the same. In
particular, we obtain
$A(\tilde{Z})=\exp\left[(\beta_0+\beta_1)\tilde{Z}^2/2\right]$ (plus
higher-order terms), i.e., the coupling function coincides with that
of a single-scalar theory with coupling constant
$\beta=\beta_0+\beta_1$.  Thus, as long as $|Z|\ll\gothr$, the theory
with $\alpha=0$ is equivalent to a ST theory with one scalar and
coupling $\beta=\beta_0+\beta_1$ (or $\beta=\beta_0-\beta_1$, in which
case only $W={\rm Im}[\psi]$ scalarizes).
Clearly, this proof also includes the limit $\gothr\to\infty$, where
the solutions reduce exactly to those of a single-scalar theory
with the identification $\beta\equiv\beta_0+\beta_1$.

When the condition $|Z|\ll\gothr$ is not fulfilled, the theory is still equivalent to a ST theory
with one scalar field,
but the form of the conformal factor $A$ changes. These theories only
differ by higher-order terms in the series
expansion~\eqref{eq:A_expansion}, \eqref{eq:psi}, which are negligible
at the onset of the scalarization.

\begin{figure}
\includegraphics[height=150pt,clip=true]{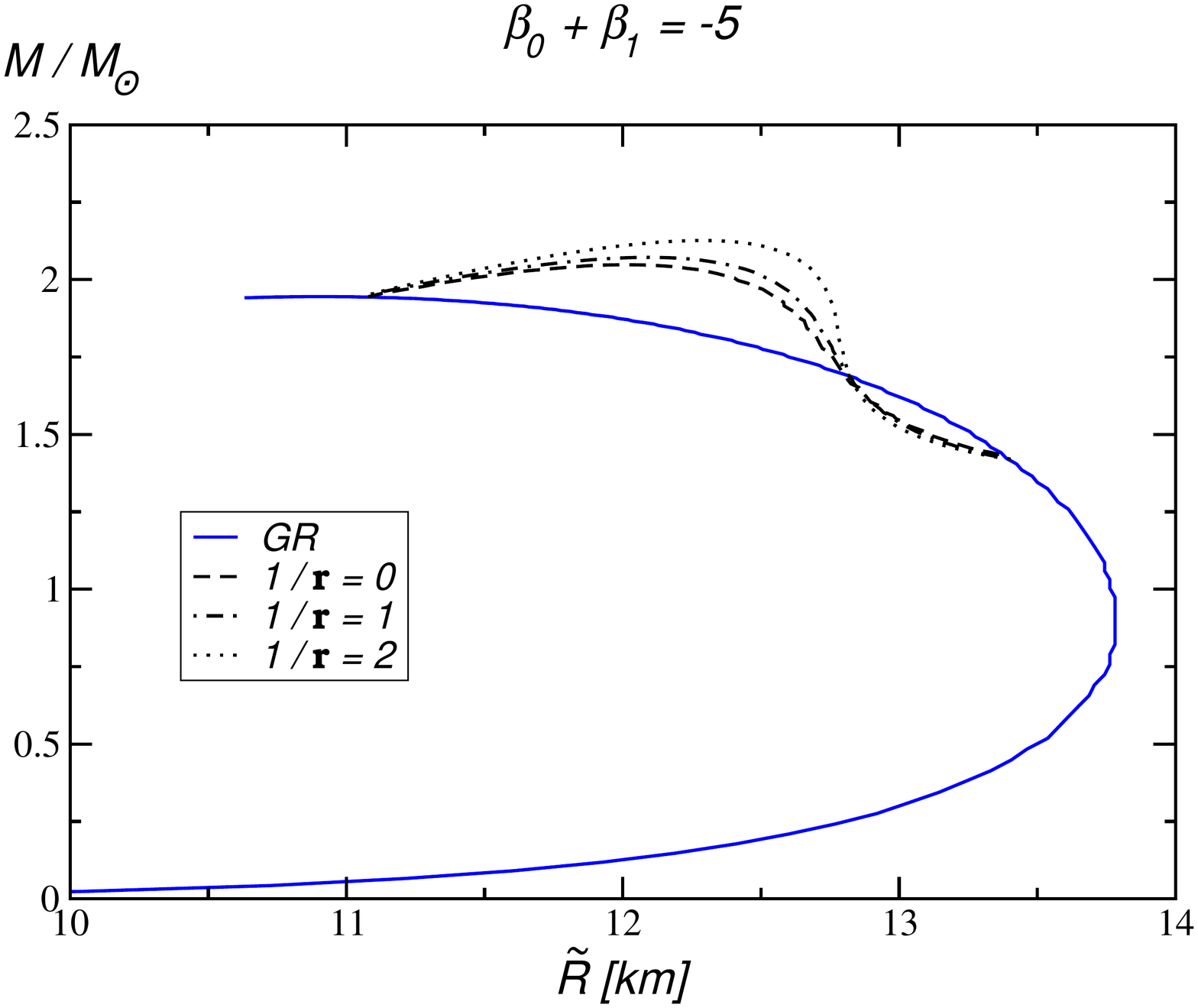}
\includegraphics[height=150pt,clip=true]{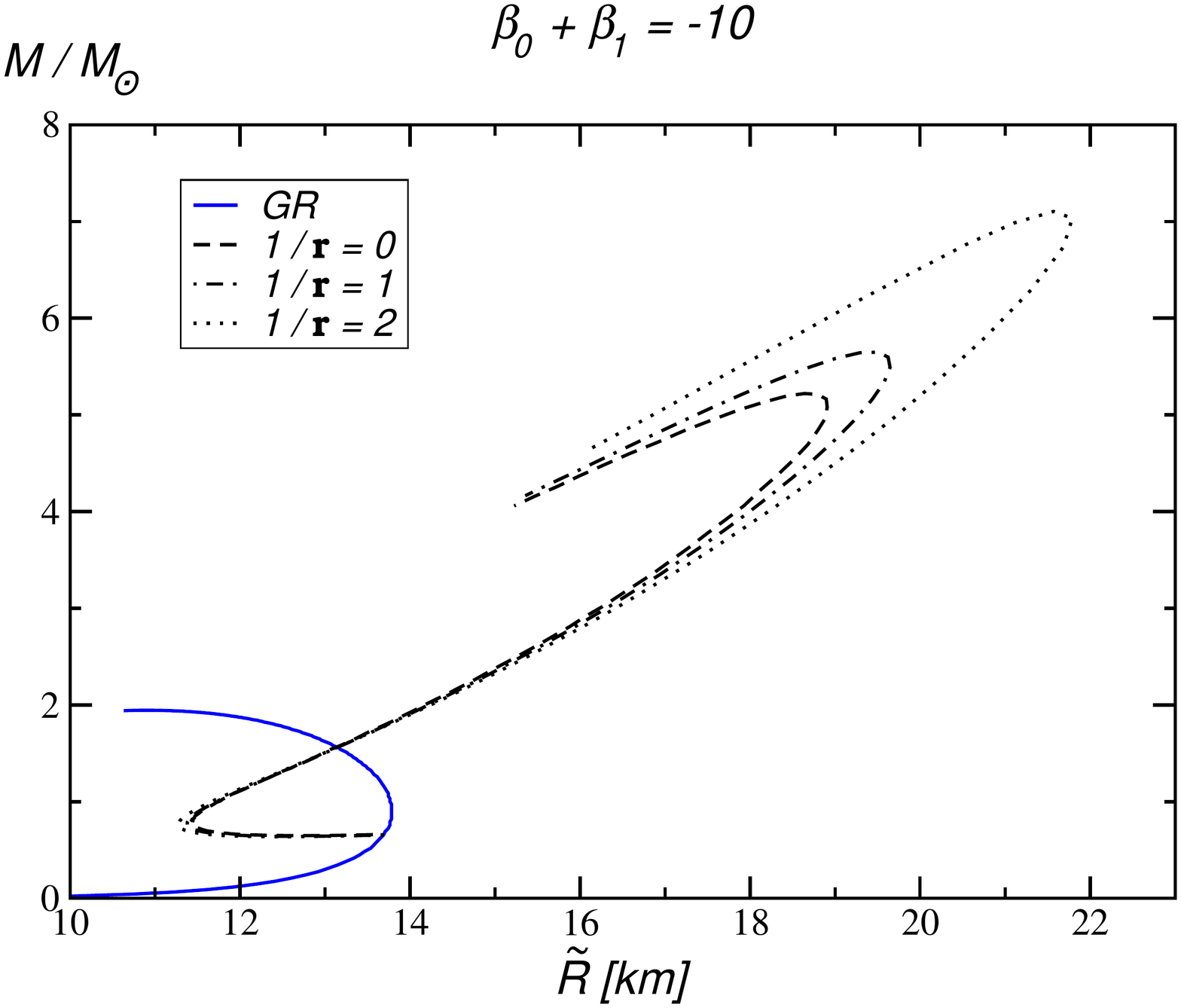}
\caption{{\it Mass-radius relations in the full TMS theory}.
  Analogous to the left panel of Fig.~\ref{fig:u_1_mass_radius} for
  three values of the curvature radius of the target metric
  ($\radcurv = \infty$, $\radcurv = 1$ and $\radcurv = 0.5$),
  $\beta_0+\beta_1=-5$ (left panel) and $\beta_0+\beta_1 = -10$ (right
  panel). Here we only consider models where ${\rm Re}[\psi]\neq 0$.
  The gravitational mass $M$ is shown as a function of the
  Jordan-frame radius $\tilde R$. For comparison, we include in both
  panels the GR curve. Note the different axis ranges in the two
  panels. When $\radcurv\to\infty$, the theory reduces to a ST theory
  with one scalar and effective coupling $\beta=\beta_0+\beta_1$, and
  the observational constraint $\beta_0+\beta_1\gtrsim-4.5$ is in
  place~\cite{Freire:2012mg}. However, such lower bound might be less
  stringent when $\radcurv$ is finite.} \label{fig:nonu_1_mass_radius}
\end{figure}

In Fig.~\ref{fig:nonu_1_mass_radius} we show the mass-radius relation
of scalarized neutron star solutions in the non-$O(2)$ symmetric theory for
different values of $\gothr$ and $\beta_0+\beta_1$. When the coupling
is large, we observe that the solutions can differ dramatically from
their GR counterpart.

%
%

\paragraph{Case $\alpha\neq0$: multi-scalarization}
When $\alpha\neq0$, GR configurations are not solutions of the field
equations. In particular, a constant (or vanishing) scalar field does
not satisfy Eq.~\eqref{fieldeqph} when $T\neq 0$. Therefore it is not
surprising that when $\alpha\neq0$ we can find solutions with two
non-trivial scalar profiles even when $\beta_0=\beta_1=0$. A more
interesting question is whether there are stellar configurations in
which both scalar fields have a large amplitude. As we have seen,
these ``biscalarized'' solutions are absent in the $\alpha=0$
case. Here we present preliminary results that demonstrate the
existence of interesting biscalarized solutions as long as
$\alpha \neq 0$.

For concreteness we set $|\alpha|=10^{-3}$: such a small value of
$|\alpha|$ satisfies the experimental bounds discussed
in~\ref{app:constraints} (but we have also studied the case where
$|\alpha|=10^{-4}$, with qualitatively similar results).

\begin{figure}
  \includegraphics[height=300pt,clip=true]{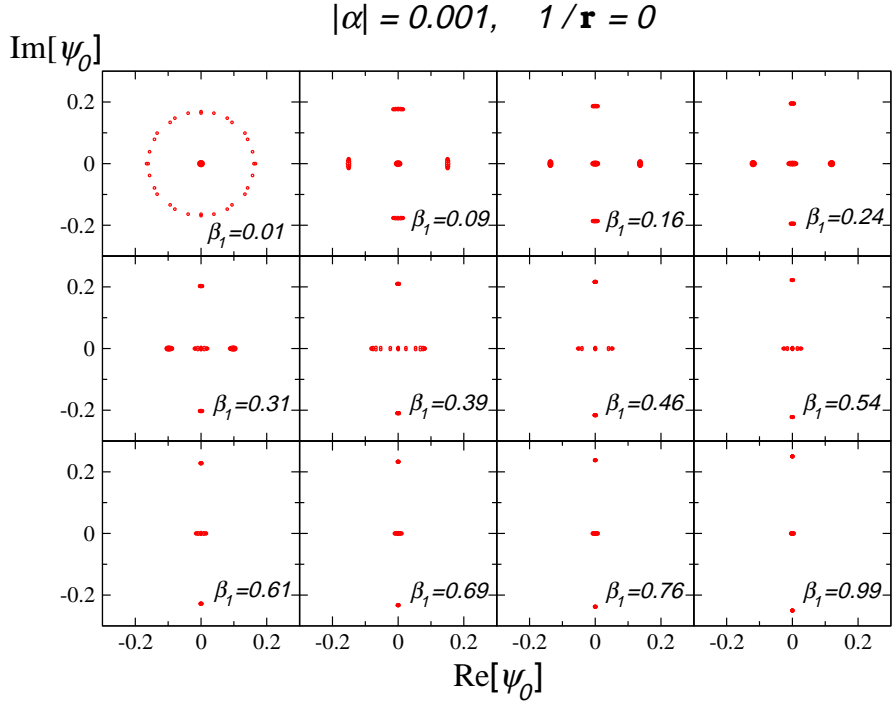}
  \caption{
          {\it Scalar field amplitudes in the full TMS theory - I.}
          Scalar field amplitude at the stellar center $\psi_0$ for
           stellar models with
           $\beta_0=-5$, $|\alpha|=0.001$ and fixed baryon mass
           $M_B = 1.8~M_{\odot}$. The different panels show the solutions
           found for different values of $\beta_1$ as indicated in each
           panel. In each case, we vary the phase of $\alpha$ from $0$ to
           $2\pi$ in steps of $\pi/6$. In contrast to the $\alpha=0$
           case in Fig.~\ref{fig:u_1_break}, the breaking of the
           $O(2)$ symmetry occurs gradually as $\beta_1$ is increased
           away from $0$.
          }
  \label{fig:bm5_alpha0001_iR0}
\end{figure}

\begin{figure}
  \centering
  \includegraphics[height=160pt,trim=0 250 0 0, clip]{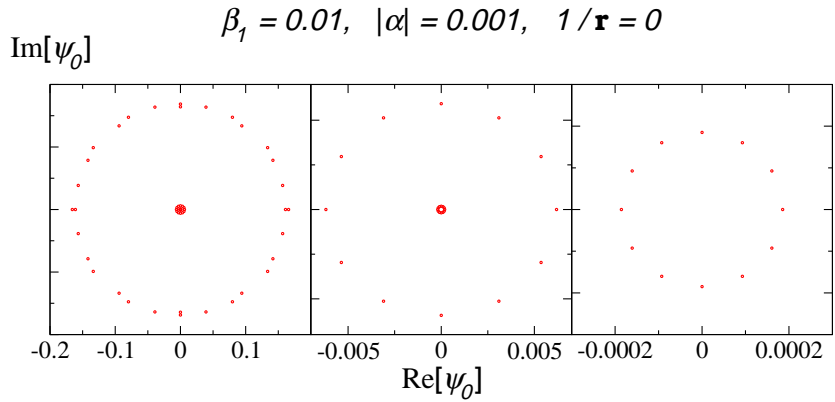}
  \caption{
          {\it Scalar field amplitudes in the full TMS theory - II.}
          The data of the upper left panel of Fig.~\ref{fig:bm5_alpha0001_iR0}
           are shown on different scales to resolve the fine structure
           of the solutions in the $\psi_0$ plane. In each panel
           the vertical extent is equal to the horizontal.}
  \label{fig:bm5_alpha0001_iR0_zoom}
\end{figure}

In this preliminary study we vary $\arg\alpha$ in the range
$[0,\,2\pi]$ in steps of $\pi/6$ and we set $1/\gothr=0$ (i.e., we
consider a flat target space).  A finite target-space curvature
$\gothr$ does not change the picture qualitatively; a more
detailed analysis will be presented elsewhere~\cite{elsewhere}.
Our search yields several models with non-zero scalar field, as shown
in Figs.~\ref{fig:bm5_alpha0001_iR0}
and~\ref{fig:bm5_alpha0001_iR0_zoom}, where dots denote the real and
imaginary parts of the central value of the scalar field $\psi_0$ for
which solutions were found.

For the time being, we wish to remark two very interesting (and
perhaps unexpected) features of these biscalarized solutions: 

\begin{itemize}
\item[1)] Figure~\ref{fig:bm5_alpha0001_iR0} shows that the solutions
  are at least approximately $O(2)$ symmetric when
  $\beta_1\sim |\alpha|$, and the $O(2)$ symmetry is broken (the
  solution ``circles'' turn into ``crosses'') when
  $\beta_1\gg |\alpha|$. The cross-like shape of the scalarized
  solutions in the ${\rm Re}[\psi_0]$, ${\rm Im}[\psi_0]$ plane
  collapses towards a set of solutions on the vertical line
  ${\rm Re}[\psi_0]=0$ for the larger values of $\beta_1$ (bottom
  panels in Fig.~\ref{fig:bm5_alpha0001_iR0}).  This behavior can be
  interpreted as an approximation to the spontaneous scalarization for
  the case $\alpha=0$ displayed in Fig.~\ref{fig:scalarization}, and
  discussed in the text around Eqs.~(\ref{eq:Kscal}) and
  (\ref{eq:Zscal}). There we observed that spontaneous scalarization
  of ${\rm Re}[\psi]$ occurs (in analogy with the single-field case)
  if $\beta_0 + \beta_1 \lesssim -4.35$, and scalarized models with a
  large imaginary part ${\rm Im}[\psi]$ exist if
  $\beta_0 - \beta_1 \lesssim -4.35$. The biscalarized models in
  Fig.~\ref{fig:scalarization} have been calculated for fixed
  $\beta_0=-5$. For $\beta_1 \gtrsim 0.65$ we therefore enter the
  regime where $\beta_0+\beta_1 \gtrsim -4.35$, and we no longer
  expect to find models with strongly scalarized ${\rm Re}[\psi]$. The
  condition $\beta_0-\beta_1 \lesssim -4.35$ for scalarization of
  ${\rm Im}[\psi]$, however, remains satisfied, so that scalarized
  models should cluster close to the ${\rm Re}[\psi_0]$--axis. This is
  indeed observed in the bottom panels of
  Fig.~\ref{fig:bm5_alpha0001_iR0}.  Note that in this case the
  condition $\beta_1 \gtrsim 0.65 \gg |\alpha| = 10^{-3}$ is
  satisfied, in close correspondence to the case $\alpha=0$ of
  Fig.~\ref {fig:scalarization}.

\item[2)] Figure~\ref{fig:bm5_alpha0001_iR0_zoom} (which is a
  ``zoom-in'' on the top-left panel of
  Figure~\ref{fig:bm5_alpha0001_iR0}) indicates additional fine
  structure in the space of solutions, with at least three different
  families of scalarized solutions having remarkably different values
  of the scalar field (and therefore of the scalar charge).

\end{itemize}

When $\gothr\to\infty$, binary pulsar observations in the
single-scalar case would impose a constraint equivalent to
$\beta_0+\beta_1\gtrsim-4.5$~\cite{Freire:2012mg}.  More work is
required to clarify whether a similar constraint is in place also for
multiple scalars. 
Preliminary calculations indicate that the target-space curvature
should affect the energy flux from compact binaries at high
post-Newtonian order.
However, it is unclear whether the formalism of \cite{Damour:1992we} 
for describing orbital binary dynamics is applicable to the theory studied 
in this paper, due to the discontinuity at $\alpha = 0$. 
Furthermore, for multiple scalars, it is possible
that some combination of $\beta_0$ and $\beta_1$ other than their sum
may be constrained by binary pulsar observations.
A detailed answer to this question requires two theoretical
developments that are currently missing: (1) the investigation of
stellar structure in generic TMS theories to understand the effect of
the theory parameters on stellar properties, and (2) an implementation
of these stellar structure calculations in flux formulas similar to
those derived in \cite{Damour:1992we} (or generalizations
thereof). These are important tasks that should be addressed in future
work.
In Fig.~\ref{fig:nonu_1_mass_radius} we adopt an agnostic point of
view and use large values of $\beta_0+\beta_1$ in order to illustrate
the effect of scalarization in TMS theory in some extreme cases.
The phenomenological implications and the stability of biscalarized
stellar models will be discussed elsewhere~\cite{elsewhere}.

\section{3+1 formulation of the field equations for numerical
  relativity}
\label{sec:3p1formulation}
Studies of the strong-field dynamics of compact stars and black holes,
whether isolated or in binary systems, require the fully non-linear
theory without any symmetry assumptions. Such studies can now be
carried out using numerical relativity techniques, and they have
already led to new insights into the behavior of ST theories. For
example, numerical simulations of neutron star binaries in
single-scalar theories have identified a new phenomenon (``dynamical
scalarization'') occurring in the late stages of the inspiral, just
before merger~\cite{Barausse:2012da,Shibata:2013pra,Ponce:2014hha}.
Similarly, studies of equilibrium sequences of neutron star binaries
have shown that dynamical scalarization can lead to a reduction of the
number of gravitational-wave cycles with respect to GR~\cite{Taniguchi:2014fqa}.
Scalar radiation has also been identified in the inspiral of black
hole binaries when the binary is embedded in a non-trivial scalar
background gradient~\cite{Healy:2011ef,Berti:2013gfa}. All numerical
studies rely on a formulation of the theory suitable for numerical
implementation, which is most commonly achieved in terms of a
space-time (or 3+1) decomposition of the field equations.  Here we
present the 3+1 formulation of the field equations for general
multi-scalar theories.
The work presented in
this section is a prerequisite for future numerical studies of
multi-scalar theories, and we hope that it will motivate other
researchers to investigate this interesting, unexplored topic.

\subsection{Action, stress-energy tensor and field equations}
We consider a multi-scalar theory described by the
action~\eqref{eq:action_ef}:
%
\begin{equation}
S=\frac{1}{4\pi G_\star}\int d^4x\sqrt{-g}\left[\frac{R}{4}-\frac{1}{2}g^{\mu\nu}\gamma_{AB}(\varphi)\partial_\mu\vp^A
  \partial_\nu\vp^B-V(\varphi)\right]+S_{\rm m}[A^{2}(\vp)g_{\mu \nu} ; \Psi] 
\,.
\label{action2}
\end{equation}

For computational purposes it is useful to consider the scalar fields themselves as ordinary matter, described by the
stress-energy tensor $T^{(\vp)}_{\mu\nu}$ defined in Eq.~\eqref{eq:Tphi}:
\begin{equation}
8\pi G_\star T^{(\vp)}_{\mu\nu}=2\gamma_{AB}\left(\partial_\mu\varphi^A\partial_\nu\varphi^B-\frac{1}{2}g_{\mu\nu}
g^{\alpha\beta}\partial_\alpha\varphi^A\partial_\beta\varphi^B\right)-2g_{\mu\nu} V \,,\label{tmunu}
\end{equation}
while $T_{\mu\nu}$, defined in Eq.~\eqref{eq:T}, is associated to the fields $\Psi$ (for instance, the fluid composing a
neutron star). The total stress-energy tensor, then, is $T_{\mu\nu}+T^{(\vp)}_{\mu\nu}$. This allows us to use the $3+1$
decomposition of Einstein's equations given in~\cite{2008itnr.book.....A}, with the replacement $T_{\mu\nu}\rightarrow
T_{\mu\nu}+T^{(\vp)}_{\mu\nu}$. Since
%
\begin{align}
\frac{16\pi G_\star}{\sqrt{-g}}\frac{\delta S}{\delta g^{\mu\nu}}=&
R_{\mu\nu}-2\gamma_{AB}\partial_\mu\varphi^A\partial_\nu\varphi^B-\frac{1}{2}g_{\mu\nu}
\left(R-2\gamma_{AB}g^{\alpha\beta}\partial_\alpha\varphi^A\partial_\beta\varphi^B-4V\right)\nonumber\\
&-8\pi G_\star T_{\mu\nu}=0\,,
\label{einsteq}
\end{align}
Einstein's equations have the form $R_{\mu\nu}-(1/2)g_{\mu\nu}R=8\pi \bareG
(T_{\mu\nu}+T^{(\vp)}_{\mu\n})$.
%
%
The trace of Eq.~(\ref{einsteq}) yields
\begin{equation}
-R+2\gamma_{AB}g^{\alpha\beta}\partial_\alpha\varphi^A\partial_\beta\varphi^B+8V-8\pi G_\star T=0\,,
\end{equation}
where $T=g^{\mu\nu}T_{\mu\nu}$, and therefore we have
\begin{equation}
  R_{\mu\nu}-2\gamma_{AB}\partial_\mu\varphi^A\partial_\nu\varphi^B-2g_{\mu\nu}V-8\pi G_\star\left(T_{\mu\nu}-\frac{1}{2}
  g_{\mu\nu}T\right)=0\,.
\end{equation}

By varying the action (\ref{action2}) with respect to $\varphi^A$ one
gets the scalar field equation
\begin{align}
  4\pi G_\star  \frac{\gamma^{AB}}{\sqrt{-g}}\frac{\delta S}{\delta\varphi^B} =&
  \Box\varphi^A+\gamma^A_{CD}g^{\mu\nu}\partial_\mu\varphi^C\partial_\nu\varphi^D
-\gamma^{AB}\frac{\partial V}{\partial\varphi^B}\nonumber\\
&+4\pi G_\star\gamma^{AB}\frac{\partial \log A}{\partial\vp^B}T=0\,,
\label{scaleq}
\end{align}
where $\gamma^A_{CD}$ are the Christoffel symbols on the target space.

\subsection{$3+1$ decomposition}
As discussed in Ref.~\cite{Berti:2013gfa} (see also
\cite{2008itnr.book.....A}), we consider a slicing of the spacetime in a set of surfaces
$\Sigma$. We introduce the normal $n_\mu$ to those surfaces and the projector
\begin{equation}
h_{\mu\nu}=g_{\mu\nu}+n_\mu n_\nu\,,\label{defproj}
\end{equation}
and write the metric in the form $(\mu,\nu=0,\dots,3;~i,j=1,2,3)$
\begin{equation}
ds^2=g_{\mu\nu}dx^\mu dx^\nu=-\alpha^2dt^2+h_{ij}(dx^i+\beta^idt)
(dx^j+\beta^jdt)\,,
\end{equation}
where $\alpha$, $\beta^i$ and $h_{ij}$ are the lapse, the shift, and
the metric on $\Sigma$, respectively. We remark that in these
coordinates $n_i=0$, therefore the $3$-dimensional metric coincides
with the projector (\ref{defproj}) restricted to the spatial
indices. It is also worth noting that $h^{0\mu}=0$, and
$g^{\mu\nu}h_{\mu\nu}=3$.

We define the covariant derivative on $\Sigma$ as
$D_i \equiv h_i{}^\alpha\nabla_\alpha$.
Since $\partial_t=\alpha n+\beta$, the Lie derivative with respect
to $n$ is ${\cal L}_n=(\partial_t-{\cal L}_\beta)/\alpha$.
The extrinsic curvature is defined as
\begin{equation}
K_{\mu\nu} \equiv -h_\mu{}^{\sigma}\nabla_\sigma n_\nu\,.
\end{equation}
Its contravariant form is purely spatial, i.e., $K^{0\mu}=0$. The
extrinsic curvature satisfies the relation
$K_{ij}={\cal L}_n h_{ij}/2$, so the evolution equation for the metric
reads
\begin{equation}
{\cal L}_n h_{ij}=-2K_{ij}\,.
\end{equation}
Other useful relations are~\cite{Salgado:2005hx,2008itnr.book.....A}
\begin{equation}
\nabla^\mu n_\mu=-K\,,\quad
n^\mu\nabla_\mu n^\nu=D^\nu(\ln\alpha)\,,
\end{equation}
where we defined $K \equiv g_{\mu\nu}K^{\mu\nu}$.

In the same way, we can define the curvature of each of the scalar fields as
$K_\varphi^A \equiv -{\cal L}_n\varphi^A/2$. Consequently, the evolution equation for the scalar fields reads
\begin{equation}
{\cal L}_n\varphi^A=-2K_\varphi^A\,,
\end{equation}
where we note that ${\cal L}_n\varphi^A=n^\mu\varphi^A_\mu$.

It will also be useful to decompose the quantity
$g^{\alpha\beta}\partial_\alpha\varphi^A\partial_\beta\varphi^B$ as follows:
\begin{equation}
g^{\alpha\beta}\partial_\alpha\varphi^A\partial_\beta\varphi^B=(h^{\alpha\beta}-n^\alpha n^\beta)\partial_\alpha
\varphi^A\partial_\beta\varphi^B=D^i\varphi^A D_i\varphi^B-4K_\varphi^AK_\varphi^B\,.
\end{equation}

\subsubsection{Einstein's equations}
The $3+1$ decomposition of Einstein's equations with matter is given e.g.  in Eqs.~(2.4.6), (2.4.9) and (2.5.6)
of~\cite{2008itnr.book.....A}; in those equations the matter terms are expressed in terms of the quantities $\rho=n^\mu
n^\nu T_{\mu\nu}$, $j^i=-h^{i\mu}n^\nu T_{\mu\nu}$, and $S_{ij}=h_i^{~\alpha} h_j^{~\beta}T_{\alpha\beta}$. We
simply replace in those equations $T_{\mu\nu}\rightarrow T_{\mu\nu}+T^{(\vp)}_{\mu\nu}$, where the explicit expression
of $T^{(\vp)}_{\mu\nu}$ is given in Eq.~\eqref{tmunu}, i.e., we replace $\rho\rightarrow\rho+\rho^{(\vp)}$,
$j^i\rightarrow j^i+j^{i(\vp)}$, $S_{ij}\rightarrow S_{ij}+S_{ij}^{(\vp)}$, where:
\begin{align}
8\pi G_\star\rho^{(\vp)}
&=\gamma_{AB}\left[D^i\varphi^AD_i\varphi^B+4K_\varphi^AK_\varphi^B\right]+2V\,,\\
8\pi G_\star j^{i(\vp)} &=-2\gamma_{AB}D^i\varphi^A(-2K_\varphi^B)=
4\gamma_{AB}D^i\varphi^AK_\varphi^B\,,\\
8\pi G_\star S_{ij}^{(\vp)} &=2\gamma_{AB}\left[D_i\varphi^A D_j\varphi^B+2 h_{ij}K_\varphi^AK_\varphi^B
  -\frac{1}{2} h_{ij}D_i\varphi^A D^i\varphi^B\right]-2h_{ij}V\,.
\end{align}
We also have:
\begin{equation}
4\pi G_\star\left[(S^{(\vp)}-\rho^{(\vp)})h_{ij}-2S^{(\vp)}_{ij}\right] =
-2\gamma_{AB}D_i\varphi^AD_j\varphi^B-2h_{ij}V\,.
\end{equation}
Then Eqs.~(2.4.6), (2.4.9) and (2.5.6) of~\cite{2008itnr.book.....A} give:
\begin{align}
^{(3)}R+K^2-K_{\mu\nu}K^{\mu\nu} =& 16\pi G_\star\rho, \nonumber\\
&+ 2\gamma_{AB}\left[D^i\varphi^AD_i\varphi^B+4K_\varphi^AK_\varphi^B\right]
+4V, \\
D_j(K^{ij}-h^{ij}K) =& 8\pi G_\star j^i+4\gamma_{AB}D^i\varphi^AK_\varphi^B
\end{align}
and
\begin{align}
{\cal L}_nK_{ij} =& -D_iD_j(\ln\alpha)+~^{(3)}R_{ij}+KK_{ij}-2K_{ik}K^k_{~j}+4\pi G_{\star}\left[h_{ij}(S-\rho)-2S_{ij}\right] \nonumber\\
&-2\gamma_{AB}D_i\varphi^AD_j\varphi^B-2h_{ij}V\,, \nonumber \\
\end{align}
where $^{(3)}R_{ij}$ and $^{(3)}R$ are the Ricci tensor and the Ricci
scalar of the metric $h_{ij}$, respectively.

\subsubsection{Scalar field equation}
To decompose the scalar equation (\ref{scaleq}), i.e.
\begin{equation}
\Box\varphi^A+\gamma^A_{CD}g^{\mu\nu}
\partial_\mu\varphi^C\partial_\nu\varphi^D -\gamma^{AB}\frac{\partial V}{\partial\varphi^B}
+4\pi G_\star\gamma^{AB}\frac{\partial\log A}{\partial\vp^B}T=0,
\end{equation}
we start by considering the first term, $\Box\varphi^A$ (the single-scalar
case was discussed in~\cite{Salgado:2005hx,Salgado:2008xh}). We have:
\begin{align}
\Box\varphi^A &=\nabla_\sigma(g^{\sigma\rho}\nabla_\rho\varphi^A) \nonumber \\
&= \nabla_\sigma
\left[(-n^\sigma n^\rho+h^{\sigma\rho})\nabla_\rho\varphi^A\right]
=\nabla_\sigma[2n^\sigma K_\varphi^a+D^\sigma\varphi^A]\nonumber\\
&=2{\cal L}_nK_\varphi^A-2KK_\varphi^A+D_iD^i\varphi^A+D_\rho(\ln\alpha)D^\rho\varphi^A\,.
\end{align}
Then, since $T=S-\rho$, the scalar field equation takes the form
\begin{align}
{\cal L}_nK_\varphi^A &= KK_\varphi^A-\frac{1}{2}D_iD^i\varphi^A-\frac{1}{2}D^i\varphi^AD_i(\ln\alpha)
\nonumber\\
&-\frac{1}{2}\gamma^A_{CD}\left(D^i\varphi^CD_i\varphi^D-4K_\varphi^CK_\varphi^D\right)
+\frac{1}{2}\gamma^{AB}\frac{\partial V}{\partial\varphi^B}\nonumber\\
&-2\pi G_\star\gamma^{AB}\frac{\partial\log A}{\partial\vp^B}(S-\rho)\,.
\end{align}

\subsection{$3+1$ equations for $2$-sphere and $2$-hyperboloid}
Let us now specialize to scalar fields living in a two-dimensional
target space with maximal symmetry and positive or negative curvature,
i.e.~a sphere or hyperboloid. For simplicity we also assume a
vanishing potential ($V\equiv 0$). As discussed
in~\ref{app:targetspaces}, the sphere and hyperboloid are both
described in stereographic coordinates by the metric
\begin{equation}
\gamma_{AB}=F\left(\begin{array}{cc}1&0\\0&1\\
\end{array}\right),
\end{equation}
with
\begin{equation}
F(Z,W)\equiv \frac{\radcurv^4}{\left[(Z^2+W^2)/4+\gothr^2 \right]^2}\,.
\end{equation}
Here $\varphi^A=(Z,W)$, and $\gothr^2$ is positive (negative) for the
sphere (hyperboloid).  The Christoffel symbols (see
\ref{app:targetspaces}) are
\begin{equation}
\gamma^Z_{AB}=\frac{1}{Z^2+W^2+\gothr^2}\left(\begin{array}{cc}-2Z&-2W\\-2W&2Z\\\end{array}\right)\,,\quad
\gamma^W_{AB}=\frac{1}{Z^2+W^2+\gothr^2}\left(\begin{array}{cc}2W&-2Z\\-2Z&-2W\\\end{array}\right)\,, \nonumber \\
\end{equation}
and Einstein's equations can be written as
\begin{align}
^{(3)}R+K^2-K_{\mu\nu}K^{\mu\nu} &=
2\gamma_{AB}\left[D^i\varphi^AD_i\varphi^B+4K_\varphi^AK_\varphi^B\right]+16\pi G_\star\rho
\nonumber\\
&= 2F \left[D^iZD_iZ+D^iWD_iW+4(K_Z^2+K_W^2)\right]+16\pi G_\star\rho\,,\\
D_j(K^{ij}-h^{ij}K) &=4\gamma_{AB}D^i\varphi^AK_\varphi^B +8\pi G_\star j^i\nonumber \\
&=4F(D^iZK_Z+D^iWK_W)+8\pi G_\star j^i\,, \\
{\cal L}_nK_{ij}=&-D_iD_j(\ln\alpha)+~^{(3)}R_{ij}+KK_{ij}-2K_{ik}K^k_{~j}\nonumber\\
&-2\gamma_{AB}D_i\varphi^AD_j\varphi^B +4\pi G_\star\left[h_{ij}(S-\rho)-2S_{ij}\right]
\nonumber\\
=&-D_iD_j(\ln\alpha)+~^{(3)}R_{ij}+KK_{ij}-2K_{ik}K^k_{~j} \nonumber\\
&-2F(D_iZD_jZ+D_iWD_jW)+4\pi G_\star\left[h_{ij}(S-\rho)-2S_{ij}\right]\,.
\end{align}
Finally, the scalar field equations are
\begin{align}
{\cal L}_nK_Z &= KK_Z-\frac{1}{2}D_iD^iZ-\frac{1}{2}D^iZD_i(\ln\alpha) -\frac{1}{2}\gamma^Z_{CD}(D^i\varphi^CD_i\varphi^D
-4K_\varphi^CK_\varphi^D)\nonumber\\
&= KK_Z-\frac{1}{2}D_iD^iZ-\frac{1}{2}D^iZD_i(\ln\alpha) +\frac{1}{Z^2+W^2
+\gothr^2}\nonumber\\
&\times\left[ \left(ZD_i ZD^i Z-ZD_i WD^iW+2W D_iZD^iW \right)\right.\nonumber\\
  &\left.-4\left(ZK_Z^2-ZK_W^2+2WK_ZK_W\right)\right]
-2\pi G_\star F^{-1}\frac{\partial\log A}{\partial Z}(S-\rho)\,,\\
{\cal L}_nK_W &= KK_W-\frac{1}{2}D_iD^iW-\frac{1}{2}D^iWD_i(\ln\alpha)-\frac{1}{2}\gamma^W_{CD}(D^i\varphi^CD_i\varphi^D
-4K_\varphi^CK_\varphi^D)\nonumber\\
&= KK_W-\frac{1}{2}D_iD^iW-\frac{1}{2}D^iWD_i(\ln\alpha)+\frac{1}{Z^2+
W^2+\gothr^2}\nonumber\\
&\times\left[\left(-WD_i ZD^i Z+WD_i WD^iW+2Z D_iZD^iW\right)\right.\nonumber\\
  &\left.-4\left(-WK_Z^2+WK_W^2+2ZK_ZK_W\right)\right]
-2\pi G_\star F^{-1}\frac{\partial\log A}{\partial W}(S-\rho)\,,\\
\end{align}
where we used the expressions of the Christoffel symbols given
in~\ref{app:targetspaces}.

\section{Discussion and conclusions}
\label{sec:conclusions}

In this paper we have barely scratched the surface of the potentially
rich phenomenology of gravitational theories with multiple scalar
fields.  Several important issues should be addressed in follow-up
work. First of all, it is important to compute experimental bounds on
the parameters $\beta_0$, $|\beta_1|$, $\gothr$, $|\alpha|$ and
$\arg\alpha - \frac{1}{2} \arg \beta_{1}$ that follow from binary
pulsar timing~\cite{Berti:2015itd,Freire:2012mg}. The quadrupole-order
scalar and tensor radiation in TMS theories was computed in
\cite{Damour:1992we}, but it is unclear whether the formalism
of~\cite{Damour:1992we} is applicable to the theories that we have
studied due to the discontinuity at $|\alpha|=0$.  In any case,
drawing exclusion diagrams in the multidimensional phase space of the
theory would require extensive stellar-structure calculations, that
will be presented in future work. Preliminary results suggest that the
target-space curvature radius $\gothr$ enters the gravitational-wave
flux (at least formally) at high post-Newtonian order, and therefore
it is quite likely that $\gothr$ will be poorly constrained by binary
pulsars. This opens the possibility of interesting new phenomenology in the
sensitivity window of advanced Earth-based gravitational-wave
detectors. Furthermore, it is unclear whether binary pulsar
observations will constrain $\beta_0$, $|\beta_1|$, or some
combination thereof, and the parameter
$\arg\alpha - \frac{1}{2} \arg \beta_{1}$ (which according to our
preliminary results plays a crucial role in ``biscalarization'') is
presently unconstrained.

In this work we have mainly presented various formal developments, but
also some new physical results, which in our opinion are
representative of the behavior of a wide class of TMS theories:
\begin{enumerate}
\item In theories with $\alpha=0$, GR solutions co-exist with
  scalarized solutions but (besides the case of $O(2)$-symmetric
  theories with $\beta_1=0$) we could not find any ``biscalarized''
  solution for any value of $\beta_0$. In other words, in this case
  either the real or the imaginary part of the complex scalar field
  scalarizes but not both, and the $O(2)$ symmetry of the $\beta_1=0$
  case is broken.
\item The $\alpha\neq0$ case is dramatically different. In this case
  -- even when $|\alpha|$ is small enough to be compatible with Solar
  System constraints -- biscalarized solutions exist, and their
  existence depends quite critically on the value of
  $\arg\alpha - \frac{1}{2} \arg \beta_{1}$. These solutions seem to
  exist quite generically in the complex-$\alpha$ plane, but their
  properties strongly depend on the values of $\beta_0$ and
  $|\beta_1|$.
\end{enumerate}
These results were obtained through extensive numerical
searches. However, given the large dimensionality of the parameter
space, we cannot exclude the existence of other solutions which were
not found in our initial searches. An approximate analytical model
which supports our results and a more detailed analysis of
biscalarization will be presented elsewhere~\cite{elsewhere}.

Some obvious extensions of the present results concern the study of
rotating neutron stars in TMS theory (generalizing
\cite{Damour:1996ke,Sotani:2010dr,Doneva:2013qva,Pani:2014jra}) and of
the universal relations valid for neutron stars in general relativity
\cite{Yagi:2013bca,Yagi:2013awa}, which may or may not hold in this
theory.
Another possible extension is to relax the assumption of a vanishing
potential in the action, i.e. to consider the multi-scalar
generalization of massive Brans-Dicke theory
\cite{Alsing:2011er,Berti:2012bp}, and investigate the effect of the
scalar field masses on the structure of scalarized neutron stars.
The use of more realistic equations of state is pivotal in confronting
TMS theory predictions on the evolution of binary pulsars with
observations.  Furthermore, we hope that the $3+1$ split worked out in
Section~\ref{sec:3p1formulation} will encourage other research groups
to perform numerical simulations in TMS theories. This may lead to
studies of phenomenological interest, such as investigations of
dynamical multi-scalarization in neutron star binaries and evolutions
of binary black-hole systems in the presence of non-trivial scalar
field backgrounds.

\section*{Acknowledgments}

M.H., H.O.S. and E.B. were funded by NSF CAREER Grant No.~PHY-1055103.
M.H. would like to acknowledge financial support from the European Research Council under the European Union’s
Seventh Framework Programme (FP7/2007-2013) / ERC grant agreement n. 306425 ``Challenging General Relativity''.
P.P. was supported by the European Community through
the Intra-European Marie Curie Contract No.~AstroGRAphy-2013-623439 and by FCT-Portugal through the project IF/00293/2013.
E.B. was supported by FCT
contract IF/00797/2014/CP1214/CT0012 under the IF2014 Programme.
D.G. is supported by
the UK STFC and the Isaac Newton Studentship of the University of
Cambridge.
U.S.~acknowledges support by the FP7-PEOPLE-2011-CIG Grant
No. 293412,
H2020 ERC Consolidator Grant 1224
Agreement No. MaGRaTh-646597,
SDSC and TACC
through XSEDE Grant No.~PHY-090003 by the NSF, Finis Terrae through
Grant No.~ICTS-CESGA-249,
STFC Roller Grant No. ST/L000636/1 and
DiRAC's Cosmos Shared Memory system through BIS Grant No.~ST/J005673/1
and STFC Grant Nos.~ST/H008586/1, ST/K00333X/1.
This work was supported by the H2020-MSCA-RISE-2015 Grant No.~StronGrHEP-690904.
We thank the support
teams of Cambridge's Cosmos system, SDSC's Trestles, TACC's Stampede
and CESGA's Finis Terrae clusters where computations have been performed.
This work was partially supported by NewCompStar (COST Action MP1304) and by
the  FP7-PEOPLE-2011-IRSES Grant No.295189 NRHEP.
\appendix

\section{Spherical and hyperboloidal target spaces}
\label{app:targetspaces}

In TMS theory~\cite{Damour:1992we}, the scalar field $\vp^A(x^{\mu})$
is an application from the space-time manifold $\mathcal{M}$ to the
target-space manifold $\mathcal{T}$. This target-space manifold is
Riemannian, and its metric is denoted by $\gamma_{AB}(\vp^C)$. The
dimensionality of $\mathcal{T}$ (i.e. the number of scalar fields) is
$N$. Since one-dimensional manifolds are necessarily flat, the
simplest non-trivial case is $N=2$. Furthermore, the simplest
two-dimensional manifolds are the maximally symmetric ones,
i.e. spherical, hyperbolic and flat spaces. In these spaces, the
curvature radius $\hat r>0$ is constant; the Ricci scalar is
${\cal R}=2/{\hat r}^2$ for spherical space, ${\cal R}=-2/{\hat r}^2$
for hyperbolic space, and ${\cal R}=0$ for flat space. For convenience
we define $\gothr=\hat r,\,{\rm i}\hat r$ for spherical and hyperbolic
spaces, respectively, so that the Ricci scalar has the form
${\cal R}=2/\gothr^2$ in both cases.

Here we derive the expression for the target-space line element
$\gamma_{AB}d\vp^A d\vp^B=2\gamma d\vp d\cvp$ in terms of the
complexified scalar field $\vp=Z+{\rm i}W$ for the spherical
and hyperbolic cases; the result is Eq.~\eqref{targmetr} in the main
text.

\subsection*{Spherical target space}

The 2-sphere can be defined from its embedding in a three-dimensional Euclidean space of coordinates $(x,y,z)$ through
the equation
\begin{equation}
x^2+y^2+z^2={\hat r}^2\,.
\end{equation}
It can be parametrized in polar coordinates, defining
$\vp^{A'}=(\Theta,\Phi)$, where:
\begin{equation}
x = \hat r\sin\Theta\cos\Phi, \qquad
y = \hat r\sin\Theta\sin\Phi, \qquad
z = \hat r\cos\Theta\,.
\end{equation}
The target-space metric in these coordinates is
\begin{equation}
  \gamma_{A'B'}=\left(\begin{array}{cc}{\hat r}^2&0\\0&{\hat r}^2\sin^2\Theta\end{array}\right)\,.
\end{equation}
This frame has undesirable features: in the flat-space limit
$\hat r\rightarrow\infty$ the metric diverges, and the kinetic term in
the action (\ref{action2}) [where for simplicity we set
$V(\Theta,\Phi)=0$]
\begin{equation}
  S=\frac{1}{4\pi G_\star}\int d^4x\sqrt{-g}\left[\frac{R}{4}-\frac{1}{2}g^{\mu\nu}{\hat r}^2
\left(\partial_\mu\Theta\partial_\nu\Theta+\sin^2\Theta\partial_\mu\Phi\partial_\nu\Phi
\right)\right]
\end{equation}
diverges as well. Moreover, the polar frame is not the most suitable
for numerical implementation, because it has coordinate singularities
at the boundary of the open intervals $0<\Theta<\pi$, $0<\Phi<2\pi$
where the coordinate system is defined.
To fix this problem we perform a stereographic projection from the
north pole of the sphere (which is the only point of the manifold not
covered by this chart) to the plane $\vp^A=(Z,W)$ tangent to the
south pole:
\begin{align}
Z &= \frac{2{\hat r}}{{\hat r}-z}x=2{\hat r}\frac{\sin\Theta}{1-\cos\Theta}\cos\Phi\,, \\
W &= \frac{2{\hat r}}{{\hat r}-z}y=2{\hat r}\frac{\sin\Theta}{1-\cos\Theta}\sin\Phi\,.
\label{stereo0}
\end{align}
With this projection the equator is mapped to the circle
$Z^2+W^2=4\hat r^2$; the upper and lower hemispheres are mapped to the
exterior and interior of this circle, respectively, and the north pole
is mapped to infinity.
Using $\sin\Theta/(1-\cos\Theta)=\cot(\Theta/2)$, the complex field
$\varphi=Z+{\rm i}W$ is written in a more compact form as
\begin{equation}
\varphi=2{\hat r}\cot(\Theta/2)e^{{\rm i}\Phi}\,.
\end{equation}
In the coordinate frame $\vp^A=(Z,W)$ the target-space metric is
\begin{equation}
\gamma_{AB}=\frac{(1-\cos\Theta)^2}{4}\delta_{AB}=
\frac{{\hat r}^4}{\left[(Z^2+W^2)/4+{\hat r}^2\right]^2}\delta_{AB};
\end{equation}
note that $(Z^2+W^2)/4+{\hat r}^2=2{\hat r^2}/(1-\cos\Theta)$. In terms of the complex field $\varphi$,
$\delta_{AB}d\vp^A d\vp^B=d\vp d\cvp$, therefore $\gamma_{AB}d\vp^A d\vp^B=2\gamma d\vp d\cvp$ (note that we
denote $\gamma=\gamma_{a\bar b}$, because $a,b$ can only take the value $1$), and
\begin{equation}
  \label{tmetricsph}
  \gamma=\frac{1}{2}\left(1+\frac{\vp\cvp}{4{\hat r}^2}\right)^{-2}\,.
\end{equation}
%
\subsection*{Hyperbolic target space}

The two-dimensional hyperbolic space of two sheets can also be defined
from its embedding into $\mathbb{R}^{1,2}$ with coordinates $(x,y,z)$
through the equation
\begin{equation}
-x^2-y^2+z^2={\hat r}^2\,.
\end{equation}
It can be parametrized in terms of $\vp^{A'}=(\Theta,\Phi)$ as
\begin{equation}
x = \hat r\sinh\Theta\cos\Phi\,, \qquad
y = \hat r\sinh\Theta\sin\Phi\,, \qquad
z = \pm\hat r\cosh\Theta\,.
\end{equation}
The target-space metric in these coordinates is
\begin{equation}
  \gamma_{A'B'}=\left(\begin{array}{cc}{\hat r}^2&0\\0&{\hat r}^2\sinh^2\Theta\end{array}\right)\,.
\end{equation}
As in the case of spherical space (see above) the metric diverges when
$\hat r\rightarrow\infty$, and the kinetic term in the action diverges
as well. Therefore we perform a stereographic projection from the
point at the top of the lower branch to the plane $\vp^A=(Z,W)$
tangent to the bottom of the upper branch.  With this projection, the
upper branch is mapped to the interior of the circle
$Z^2+W^2=4\hat r^2$, and the lower branch is mapped to the exterior of
this circle. The stereographic mapping reads
\begin{align}
Z &= \frac{2{\hat r}}{z+{\hat r}}x=2{\hat r}\frac{\sinh\Theta}{\cosh\Theta+1}\cos\Phi\,, \\
W &= \frac{2{\hat r}}{z+{\hat r}}y=2{\hat r}\frac{\sinh\Theta}{\cosh\Theta+1}\sin\Phi\,.
\label{stereo1_up}
\end{align}
for the upper branch, and
\begin{align}
Z &= \frac{2{\hat r}}{-z-{\hat r}}x=2{\hat r}\frac{\sinh\Theta}{\cosh\Theta-1}\cos\Phi\,, \\
W &= \frac{2{\hat r}}{-z-{\hat r}}y=2{\hat r}\frac{\sinh\Theta}{\cosh\Theta-1}\sin\Phi
\label{stereo1_dw}
\end{align}
for the lower branch.  The complex field $\varphi=Z+{\rm i}W$ is then
$\varphi=2{\hat r}\tanh(\Theta/2)e^{{\rm i}\Phi}$ for the upper
branch, and $\varphi=2{\hat r}\coth(\Theta/2)e^{{\rm i}\Phi}$ for the
lower branch.

In the coordinate frame $\vp^A=(Z,W)$ the target-space metric is
\begin{equation}
  \gamma_{AB}=\frac{(1\pm\cosh\Theta)^2}{4}\delta_{AB}=
\frac{{\hat r}^4}{\left[-(Z^2+W^2)/4+{\hat r}^2\right]^2}\delta_{AB}
\end{equation}
where the upper (lower) sign refers to the upper (lower) branch; note that
$-(Z^2+W^2)/4+{\hat r}^2=2{\hat r^2}/(1\pm\cos\Theta)$. In terms of
the complex field $\varphi$, then, the target-space metric is
$2\gamma d\vp d\cvp$ with
\begin{equation}
  \label{tmetrichyp}
\gamma=\frac{1}{2}\left(1-\frac{\vp\cvp}{4{\hat r}^2}\right)^{-2}\,.
\end{equation}
%
\subsection*{Field equations for two-dimensional spherical and hyperbolic spaces}

In summary, the expressions~\eqref{tmetricsph},~\eqref{tmetrichyp} for the target-space metric in the (two-dimensional) spherical
and hyperbolic cases can be written in the form of Eq.~\eqref{targmetr}, i.e.
\begin{equation}
  \label{tmetric}
\gamma=\frac{1}{2}\left(1+\frac{\vp\cvp}{4\gothr^2}\right)^{-2}\,,
\end{equation}
where $\gothr=\hat r$ for a spherical space, and $\gothr={\rm i}\hat r$ for a hyperbolic space. In the coordinate frame
$\vp^A=(Z,W)$ the target-space metric is
\begin{equation}
  \gamma_{AB}=\frac{{\gothr}^4}{\left[(Z^2+W^2)/4+{\gothr}^2\right]^2}\delta_{AB}
  \label{tmetric2}
\end{equation}
for both the spherical and hyperbolic space. Therefore, Eqs.~\eqref{tmetric},~\eqref{tmetric2} describe a spherical
space if $\gothr^2>0$, an hyperbolic space if $\gothr^2<0$.  The limit $\gothr\rightarrow\infty$ yields flat space. If
$\gothr\rightarrow\infty$ and the scalar field is restricted to real values, one recovers the single-scalar case.

The Christoffel symbols are:
\begin{equation}
\begin{array}{ccc}
\gamma^Z_{\phantom{Z}ZZ}=-\frac{2Z}{\gothr^2+Z^2+W^2}\,,&\gamma^W_{\phantom{W}ZZ}=\frac{2W}{\gothr^2+Z^2+W^2}\,,&
\gamma^Z_{\phantom{Z}ZW}=-\frac{2W}{\gothr^2+Z^2+W^2}\,,\\
&&\\
\gamma^W_{\phantom{W}WW}=-\frac{2W}{\gothr^2+Z^2+W^2}\,,&\gamma^Z_{\phantom{Z}WW}=\frac{2Z}{\gothr^2+Z^2+W^2}\,,&
\gamma^W_{\phantom{W}ZW}=-\frac{2Z}{\gothr^2+Z^2+W^2}\,.\\
\end{array}
\end{equation}
In terms of the complex field $\varphi$, writing explicitly the
indices $a,b$ in $\gamma_{a\bar b}$ (which can only take the value
$1$) we get
\begin{equation}
\gamma^{\bar c}_{\phantom{\bar c} a\bar b} = \frac{1}{2}\partial_\vp\log\gamma=-\frac{\cvp}{4\gothr^2+\cvp\vp}, \qquad
\gamma^{c}_{\phantom{c} a\bar b} = \frac{1}{2}\partial_{\cvp}\log\gamma=-\frac{\vp}{4\gothr^2+\cvp\vp}\,.
\label{chrcmpl}
\end{equation}
The Ricci tensor and Ricci scalar of the target space are
${\cal R}_{AB}=\gothr^{-2}\delta_{AB}$ and ${\cal R}=2\gothr^{-2}$,
respectively.

Replacing the expression of the metric~\eqref{tmetrichyp} and of the
Christoffel symbols~\eqref{chrcmpl} in
Eqs.~\eqref{eq:fieldeq_ef_tens_complex2} and
\eqref{eq:fieldeq_ef_sc_complex2} with $V(\vp)=0$ we find the field
equations for a maximally symmetric two-dimensional target space,
i.e. Eqs.~\eqref{fieldeqR} and~\eqref{fieldeqph}.

\section{Solar System constraints}\label{app:constraints}

The weak-field limit of TMS theories has been worked
out in~\cite{Damour:1992we}.  Specializing these results to the theory
constructed in the body of the text, and rewriting them in complex
notation, one finds that the gravitational constant measured in a
Cavendish experiment is given by
\begin{equation}
G = \bareG A^{2}_{\infty}
(1 +\bar{\kappa}_{\infty}  \kappa_{\infty} ) \,,
\label{eq:GGstar}
\end{equation}
where the subscript $\infty$ denotes evaluation at $\varphi_{\infty}=0$ and we defined the complex function $\kappa(\varphi,\bar{\varphi})$ as in Eq.~\eqref{eq:kappa_def}.
Using Eq.~\eqref{eq:A_expansion}, one finds that $\kappa_\infty=2\alpha^\ast$.

It is straightforward to show that the post-Newtonian parameter $\gamma_{\rm PPN}$ reads~\cite{Damour:1992we}
\begin{equation}
\gamma_{\rm PPN} - 1 = -\frac{2\bar{\kappa}_{\infty}
\kappa_{\infty}}
{1+ \bar{\kappa}_{\infty} \kappa_{\infty}}=
-\frac{8|\alpha^\ast|^2}{1+ 4|\alpha^\ast|^2} \,,
\end{equation}
and therefore the Cassini bound
$|\gamma_{\rm PPN}-1| < 2.3 \cdot 10^{-5}$~\cite{Bertotti:2003rm}
implies the constraint
\begin{equation}
|\alpha^\ast|^2 < 3 \cdot 10^{-6}
\label{boundalpha}
\end{equation}
on the coupling constants $\alpha^{\ast}$ and $\bar\alpha^{\ast}$ appearing in Eq.~\eqref{eq:A_expansion}. Crucially, the one above is a bound on $|\alpha^{\ast}|$, whereas $\arg\alpha^{\ast}$ is completely unconstrained in the weak-field limit.

On the other hand, the post-Newtonian parameter $\beta_{\rm PPN}$ reads~\cite{Damour:1992we}
\begin{equation}
\beta_{\rm PPN} - 1 =
\frac{\bar{\kappa}_{\infty}  \kappa_{\infty} \beta_{\infty}}
{2(1+\bar{\kappa}_{\infty}\kappa_{\infty})^{2}} \,,
\end{equation}
where the real-valued function $\beta(\varphi, \bar{\varphi})$ is defined by
\begin{equation}
\beta(\varphi, \bar{\varphi})  \equiv \frac{1}{2}
\left( 1 + \frac{\bar{\varphi} \varphi}{4\radcurv^{2}}\right)
\left( \kappa \frac{\partial}{\partial \bar{\varphi}}
+ \bar{\kappa} \frac{\partial}{\partial \varphi} \right) \log (\bar{\kappa} \kappa) \,.
\end{equation}
Using the definitions above and Eq.~\eqref{eq:A_expansion} we obtain
\begin{align}
\beta_{\rm PPN} - 1 &=\frac{{\alpha^\ast}{\alpha^\ast} {\bar\beta_1^\ast}+2 {\alpha^\ast} {\bar\alpha^\ast} {\beta_0}+{{\bar\alpha}^\ast}{{\bar\alpha}^\ast} {\beta_1^\ast}}{(1+ 4 {\alpha^\ast}{\bar\alpha^\ast})^2}
\nn
\\
&=
\frac{2|\alpha^\ast|^2}{(1+4|\alpha^\ast|^2)^2}
\biggl(
\beta_0 + |\beta_1^\ast| \cos (2 \arg \alpha^\ast - \arg \beta_1^\ast) 
\biggr) \,.
\label{eq:argalpha}
\end{align}

Finally, the bound $|\beta_{\rm PPN} - 1| < 1.1 \cdot 10^{-4}$ coming
from the combination of Cassini and Lunar Laser Ranging
measurements~\cite{Williams:2005rv} implies a constraint on some
combination of the parameters $\beta_0$, $|\beta_1^\ast|$, and
$\arg \alpha^\ast - \frac{1}{2} \arg \beta_{1}^{\ast}$.
However, note that if $|\alpha^\ast|\to0$ the
observational constraint $|\beta_{\rm PPN} - 1|< 1.1 \cdot 10^{-4}$ is
satisfied for any value of $\beta_0$, $|\beta_1^\ast|$ and
$\arg \alpha^\ast - \frac{1}{2} \arg \beta_{1}^{\ast}$, and therefore these parameters are unconstrained
by weak-field observations in this limit.

\section{Linearized field equations and scalarization}
\label{app:linear}
Here we consider the ST theory defined by Eqs.~(\ref{Einstein_psi})
and (\ref{eq:logA}) with $\alpha=0$, which admits GR solutions with
$\psi \equiv 0$.  We will perturb these GR solutions, and linearize
the field equations in the perturbations.  This is valid when the
amplitudes of the scalar fields are small and consequently the metric
back-reaction on the scalar field can be neglected.  This
approximation is well motivated at the onset of
scalarization.

The field equations acquire a particularly simple form when linearized
to first order in $Z\equiv {\rm Re}[\psi]$ and
$W\equiv {\rm Im}[\psi]$.  In this case, the tensor field
equations~\eqref{Einstein_psi} reduce to
\begin{equation}
R_{\mu\nu} = 8\pi G_{\star}\left( T_{\mu\nu}
- \frac{1}{2}Tg_{\mu\nu} \right)\,,
\end{equation}
and therefore the background geometry to ${\cal O}(Z,W)$ is described by a GR solution. The scalar-field
equation~(\ref{eq:psi}) becomes
\begin{align}
\square Z &= -4\pi G_{\star} (\beta_0+\beta_1) T Z\,, \label{eqlinZ}\\
\square W &= -4\pi G_{\star} (\beta_0-\beta_1) T W\,, \label{eqlinW}
\end{align}
where, in this perturbative expansion, the box operator is evaluated
on the GR background solution and the trace of the perfect fluid
energy-momentum tensor $T$ attains its GR value, i.e.
$T=-(8\pi G_\star)^{-1}R=3P-\rho$.  Note that the equations for $Z$
and $W$ decouple in this limit, reducing to the same equation as in
the single-scalar case, $\square \vp = -4\pi G_{\star} \beta T \vp$,
but with effective coupling parameters $\beta =\beta_0+\beta_1$ and
$\beta = \beta_0-\beta_1$, respectively.

In the case of a single scalar, the term on the right-hand side of the scalar equation can be interpreted as an
effective mass term (cf. e.g.~\cite{Berti:2015itd})
\begin{equation}
 m^2_{\rm eff}=-4\pi G_{\star} \beta T\,.
\end{equation}
Because in typical configurations $T\sim -\rho<0$, the effective mass squared is negative when $\beta<0$.  This signals
a possible tachyonic instability which is associated with an exponentially growing mode and causes the growth of scalar
hair in a process known as {\it spontaneous scalarization}~\cite{Damour:1993hw}, as discussed in the main text. In the case of static compact stars, it turns out that this instability occurs for $\beta\lesssim -4.35$, the threshold value
depending only mildly on the equation of
state~\cite{Harada:1997mr,Novak:1998rk,Silva:2014fca}.

The same reasoning can be applied to Eqs.~\eqref{eqlinZ} and~\eqref{eqlinW}. Because the latter are completely
equivalent to two copies of a single-scalar equation, scalarization is expected whenever
\begin{equation}
 \beta_0+\beta_1\lesssim -4.35 \quad \rm{or} \quad \beta_0-\beta_1\lesssim -4.35\,.  \label{doublescalarization}
\end{equation}

Note that these conditions were derived assuming that each scalar
field acquires a non-vanishing expectation value independently and by
perturbing a static GR solution. In particular, they do not imply that
\emph{both} fields scalarize when both
conditions~\eqref{doublescalarization} are satisfied. In fact, biscalarization can be investigated in this perturbative framework by
studying the linear perturbations of (say) the scalar field $W$ on the
background of a previously scalarized solution where $Z$ has a
non-trivial profile.

  
\section*{References}


\begin{thebibliography}{10}
\expandafter\ifx\csname url\endcsname\relax
  \def\url#1{{\tt #1}}\fi
\expandafter\ifx\csname urlprefix\endcsname\relax\def\urlprefix{URL }\fi
\providecommand{\eprint}[2][]{\url{#2}}

\bibitem{Berti:2015itd}
Berti E, Barausse E, Cardoso V, Gualtieri L, Pani P {\em et~al.\/} 2015
  (\textit{Preprint} \eprint{1501.07274})

\bibitem{polchinski1998string}
Polchinski J 1998 {\em String theory\/} (Cambridge university press)

\bibitem{Duff:1994tn}
Duff M~J 1994 {Kaluza-Klein theory in perspective} {\em {The Oskar Klein
  centenary. Proceedings, Symposium, Stockholm, Sweden, September 19-21,
  1994}\/} (\textit{Preprint} \eprint{hep-th/9410046})

\bibitem{Randall:1999ee}
Randall L and Sundrum R 1999 {\em Phys. Rev. Lett.\/} {\bf 83} 3370--3373
  (\textit{Preprint} \eprint{hep-ph/9905221})

\bibitem{Randall:1999vf}
Randall L and Sundrum R 1999 {\em Phys. Rev. Lett.\/} {\bf 83} 4690--4693
  (\textit{Preprint} \eprint{hep-th/9906064})

\bibitem{Clifton:2011jh}
Clifton T, Ferreira P~G, Padilla A and Skordis C 2012 {\em Phys.Rept.\/} {\bf
  513} 1--189 (\textit{Preprint} \eprint{1106.2476})

\bibitem{Jordan:1959eg}
Jordan P 1959 {\em Z. Phys.\/} {\bf 157} 112--121

\bibitem{Fierz:1956zz}
Fierz M 1956 {\em Helv. Phys. Acta\/} {\bf 29} 128--134

\bibitem{Brans:1961sx}
Brans C and Dicke R 1961 {\em Phys.Rev.\/} {\bf 124} 925--935

\bibitem{Will:2005va}
Will C~M 2014 {\em Living Reviews in Relativity\/} {\bf 17} 4
  \urlprefix\url{http://www.livingreviews.org/lrr-2014-4}

\bibitem{Bergmann:1968ve}
Bergmann P~G 1968 {\em Int.J.Theor.Phys.\/} {\bf 1} 25--36

\bibitem{Wagoner:1970vr}
Wagoner R~V 1970 {\em Phys.Rev.\/} {\bf D1} 3209--3216

\bibitem{Damour:1992we}
Damour T and Esposito-Far{\`e}se G 1992 {\em Class. Quant. Grav.\/} {\bf 9}
  2093--2176

\bibitem{Kainulainen:2004vk}
Kainulainen K and Sunhede D 2006 {\em Phys.Rev.\/} {\bf D73} 083510
  (\textit{Preprint} \eprint{astro-ph/0412609})

\bibitem{Albrecht:2001xt}
Albrecht A, Burgess C, Ravndal F and Skordis C 2002 {\em Phys.Rev.\/} {\bf D65}
  123507 (\textit{Preprint} \eprint{astro-ph/0107573})

\bibitem{Damour:1994ya}
Damour T and Polyakov A~M 1994 {\em Gen.Rel.Grav.\/} {\bf 26} 1171--1176
  (\textit{Preprint} \eprint{gr-qc/9411069})

\bibitem{Damour:1994zq}
Damour T and Polyakov A~M 1994 {\em Nucl.Phys.\/} {\bf B423} 532--558
  (\textit{Preprint} \eprint{hep-th/9401069})

\bibitem{Chiba:1997ms}
Chiba T, Harada T and Nakao K~i 1997 {\em Prog.Theor.Phys.Suppl.\/} {\bf 128}
  335--372

\bibitem{2003sttg.book.....F}
{Fujii} Y and {Maeda} K~I 2003 {\em {The Scalar-Tensor Theory of
  Gravitation}\/} (Cambridge University Press)

\bibitem{Horbatsch:2010hj}
Horbatsch M and Burgess C 2011 {\em JCAP\/} {\bf 1108} 027 (\textit{Preprint}
  \eprint{1006.4411})

\bibitem{Sotiriou:2015lxa}
Sotiriou T~P 2014 {\em Lect. Notes Phys.\/} {\bf 892} 3--24 (\textit{Preprint}
  \eprint{1404.2955})

\bibitem{Bertotti:2003rm}
Bertotti B, Iess L and Tortora P 2003 {\em Nature\/} {\bf 425} 374

\bibitem{Damour:1993hw}
Damour T and Esposito-Far{\`e}se G 1993 {\em Phys. Rev. Lett.\/} {\bf 70}
  2220--2223

\bibitem{Freire:2012mg}
Freire P~C, Wex N, Esposito-Farese G, Verbiest J~P, Bailes M {\em et~al.\/}
  2012 {\em Mon.Not.Roy.Astron.Soc.\/} {\bf 423} 3328 (\textit{Preprint}
  \eprint{1205.1450})

\bibitem{Barausse:2012da}
Barausse E, Palenzuela C, Ponce M and Lehner L 2013 {\em Phys.Rev.\/} {\bf D87}
  081506 (\textit{Preprint} \eprint{1212.5053})

\bibitem{Palenzuela:2013hsa}
Palenzuela C, Barausse E, Ponce M and Lehner L 2014 {\em Phys.Rev.\/} {\bf D89}
  044024 (\textit{Preprint} \eprint{1310.4481})

\bibitem{Shibata:2013pra}
Shibata M, Taniguchi K, Okawa H and Buonanno A 2014 {\em Phys.Rev.\/} {\bf D89}
  084005 (\textit{Preprint} \eprint{1310.0627})

\bibitem{Sampson:2014qqa}
Sampson L, Yunes N, Cornish N, Ponce M, Barausse E {\em et~al.\/} 2014 {\em
  Phys.Rev.\/} {\bf D90} 124091 (\textit{Preprint} \eprint{1407.7038})

\bibitem{Taniguchi:2014fqa}
Taniguchi K, Shibata M and Buonanno A 2015 {\em Phys.Rev.\/} {\bf D91} 024033
  (\textit{Preprint} \eprint{1410.0738})

\bibitem{Novak:1997hw}
Novak J 1998 {\em Phys.Rev.\/} {\bf D57} 4789--4801 (\textit{Preprint}
  \eprint{gr-qc/9707041})

\bibitem{Harada:1996wt}
Harada T, Chiba T, Nakao K~i and Nakamura T 1997 {\em Phys.Rev.\/} {\bf D55}
  2024--2037 (\textit{Preprint} \eprint{gr-qc/9611031})

\bibitem{Herdeiro:2014goa}
Herdeiro C~A~R and Radu E 2014 {\em Phys.Rev.Lett.\/} {\bf 112} 221101
  (\textit{Preprint} \eprint{1403.2757})

\bibitem{Herdeiro:2015waa}
Herdeiro C~A~R and Radu E 2015 {\em Int. J. Mod. Phys. D\/} {\bf 24} 1542014
  (\textit{Preprint} \eprint{1504.08209})

\bibitem{Bekenstein:1995un}
Bekenstein J 1995 {\em Phys.Rev.\/} {\bf D51} 6608--6611

\bibitem{Sotiriou:2011dz}
Sotiriou T~P and Faraoni V 2012 {\em Phys.Rev.Lett.\/} {\bf 108} 081103
  (\textit{Preprint} \eprint{1109.6324})

\bibitem{Heusler:1995qj}
Heusler M 1995 {\em Class.Quant.Grav.\/} {\bf 12} 2021--2036 (\textit{Preprint}
  \eprint{gr-qc/9503053})

\bibitem{Chrusciel:2012jk}
Chrusciel P~T, Costa J~L and Heusler M 2012 {\em Living Rev.Rel.\/} {\bf 15} 7
  (\textit{Preprint} \eprint{1205.6112})

\bibitem{HeuslerBook}
Heusler M 1996 {\em {Black Hole Uniqueness Theorems}\/} (Cambridge:
  Cambridge University Press)

\bibitem{Sotiriou:2015pka}
Sotiriou T~P 2015  (\textit{Preprint} \eprint{1505.00248})

\bibitem{Will:1989sk}
Will C~M and Zaglauer H~W 1989 {\em Astrophys. J.\/} {\bf 346} 366

\bibitem{Mirshekari:2013vb}
Mirshekari S and Will C~M 2013 {\em Phys.Rev.\/} {\bf D87} 084070
  (\textit{Preprint} \eprint{1301.4680})

\bibitem{Yunes:2011aa}
Yunes N, Pani P and Cardoso V 2012 {\em Phys.Rev.\/} {\bf D85} 102003
  (\textit{Preprint} \eprint{1112.3351})

\bibitem{Horbatsch:2011ye}
Horbatsch M and Burgess C 2012 {\em JCAP\/} {\bf 1205} 010 (\textit{Preprint}
  \eprint{1111.4009})

\bibitem{Healy:2011ef}
Healy J, Bode T, Haas R, Pazos E, Laguna P {\em et~al.\/} 2012 {\em
  Class.Quant.Grav.\/} {\bf 29} 232002 (\textit{Preprint} \eprint{1112.3928})

\bibitem{Berti:2013gfa}
Berti E, Cardoso V, Gualtieri L, Horbatsch M and Sperhake U 2013 {\em
  Phys.Rev.\/} {\bf D87} 124020 (\textit{Preprint} \eprint{1304.2836})

\bibitem{Damour:1996ke}
Damour T and Esposito-Farese G 1996 {\em Phys.Rev.\/} {\bf D54} 1474--1491
  (\textit{Preprint} \eprint{gr-qc/9602056})

\bibitem{1973grav.book.....M}
{Misner} C~W, {Thorne} K~S and {Wheeler} J~A 1973 {\em {Gravitation}\/}
  (W. H. Freeman, San Francisco)

\bibitem{moroianu_book}
{Moroianu} A 2007 {\em {Lectures on K\"{a}hler Geometry}\/} (Cambridge
  University Press)

\bibitem{Hartle:1967he}
Hartle J~B 1967 {\em Astrophys.J.\/} {\bf 150} 1005--1029

\bibitem{Hartle:1968ht}
{Hartle} J~B and {Thorne} K~S 1968 {\em \apj\/} {\bf 153} 807

\bibitem{elsewhere}
Horbatsch M, Silva H~O, Gerosa D, Pani P, Berti E, Gualtieri L and Sperhake U
  {in preparation}

\bibitem{Ponce:2014hha}
Ponce M, Palenzuela C, Barausse E and Lehner L 2015 {\em Phys.Rev.\/} {\bf D91}
  084038 (\textit{Preprint} \eprint{1410.0638})

\bibitem{2008itnr.book.....A}
{Alcubierre} M 2008 {\em {Introduction to 3+1 Numerical Relativity}\/} (Oxford
  University Press)

\bibitem{Salgado:2005hx}
Salgado M 2006 {\em Class.Quant.Grav.\/} {\bf 23} 4719--4742 (\textit{Preprint}
  \eprint{gr-qc/0509001})

\bibitem{Salgado:2008xh}
Salgado M, Rio D~M~d, Alcubierre M and Nunez D 2008 {\em Phys. Rev.\/} {\bf
  D77} 104010 (\textit{Preprint} \eprint{0801.2372})

\bibitem{Sotani:2010dr}
Sotani H 2010 {\em Phys.Rev.\/} {\bf D81} 084006 (\textit{Preprint}
  \eprint{1003.2575})

\bibitem{Doneva:2013qva}
Doneva D~D, Yazadjiev S~S, Stergioulas N and Kokkotas K~D 2013 {\em
  Phys.Rev.\/} {\bf D88} 084060 (\textit{Preprint} \eprint{1309.0605})

\bibitem{Pani:2014jra}
Pani P and Berti E 2014 {\em Phys.Rev.\/} {\bf D90} 024025 (\textit{Preprint}
  \eprint{1405.4547})

\bibitem{Yagi:2013bca}
Yagi K and Yunes N 2013 {\em Science\/} {\bf 341} 365--368 (\textit{Preprint}
  \eprint{1302.4499})

\bibitem{Yagi:2013awa}
Yagi K and Yunes N 2013 {\em Phys.Rev.\/} {\bf D88} 023009 (\textit{Preprint}
  \eprint{1303.1528})

\bibitem{Alsing:2011er}
Alsing J, Berti E, Will C~M and Zaglauer H 2012 {\em Phys.Rev.\/} {\bf D85}
  064041 (\textit{Preprint} \eprint{1112.4903})

\bibitem{Berti:2012bp}
Berti E, Gualtieri L, Horbatsch M and Alsing J 2012 {\em Phys.Rev.\/} {\bf D85}
  122005 (\textit{Preprint} \eprint{1204.4340})

\bibitem{Williams:2005rv}
Williams J~G, Turyshev S~G and Boggs D~H 2009 {\em Int.J.Mod.Phys.\/} {\bf D18}
  1129--1175 (\textit{Preprint} \eprint{gr-qc/0507083})

\bibitem{Harada:1997mr}
Harada T 1997 {\em Prog.Theor.Phys.\/} {\bf 98} 359--379 (\textit{Preprint}
  \eprint{gr-qc/9706014})

\bibitem{Novak:1998rk}
Novak J 1998 {\em Phys.Rev.\/} {\bf D58} 064019 (\textit{Preprint}
  \eprint{gr-qc/9806022})

\bibitem{Silva:2014fca}
Silva H~O, Macedo C~F~B, Berti E and Crispino L~C~B 2015 {\em Class. Quant.
  Grav.\/} {\bf 32} 145008 (\textit{Preprint} \eprint{1411.6286})

\end{thebibliography}

\providecommand{\newblock}{}

\end{document}